\documentclass{aa} 
\usepackage{lscape}
\usepackage{graphicx}
\usepackage{txfonts}
\begin{document}
%
   \title{Black-hole masses of type~1 AGN in the {\it XMM-Newton} bright 
serendipitous survey\thanks{Based on 
observations collected at the Telescopio Nazionale Galileo (TNG) and
at the European
Southern Observatory (ESO), La Silla, Chile and on observations
obtained with {\it XMM-Newton}, an ESA science mission
with instruments and contributions directly funded by ESA Member States and 
the USA (NASA)}}

   \subtitle{}

   \author{A. Caccianiga\inst{1},          
           R. Fanali\inst{1,2},
           P. Severgnini\inst{1},
	   R. Della Ceca\inst{1},
           E. Marchese\inst{1},
           S. Mateos\inst{3,4}
          }
   \offprints{A. Caccianiga}
   \institute{INAF - Osservatorio Astronomico di Brera, via Brera 28, 
 I-20121 Milan, Italy\\
              \email{alessandro.caccianiga@brera.inaf.it}
         \and
Dipartimento di Fisica, Universit\'a degli Studi di Milano-Bicocca,
Piazza Della Scienza 3, 20126 Milano, Italy
\and
Instituto de F\'\i sica de Cantabria (CSIC-UC), Avenida de los
Castros, 39005 Santander, Spain
\and
     X-ray \& Observational Astronomy Group, Department of Physics and 
Astronomy, 
     Leicester University, Leicester LE1 7RH, UK
            }
   \date{}

 
\abstract
{}
{We derive masses of the central supermassive black hole (SMBH) and 
accretion rates for 154 type~1 AGN belonging to a well-defined X-ray-selected 
sample, the XMM-{\it Newton} Serendipitous Sample (XBS).}
{We used the most recent ``single-epoch'' relations, based on  
H$\beta$ and MgII$\lambda$2798\AA\ emission lines, to derive the SMBH  
masses. We then used the bolometric luminosities, computed
on the basis of an SED-fitting procedure, to calculate the accretion
rates, both absolute and normalized to the Eddington luminosity (Eddington 
ratio).}
{The selected AGNs cover a range of masses from 10$^{7}$ to 10$^{10}$ 
M$_{\sun}$ with a peak around 8$\times$10$^{8}$ M$_{\sun}$ and a range of 
accretion rates from 0.01 to $\sim$50 M$_{\sun}$/year (assuming an efficiency 
of 0.1), with a peak at $\sim$1 M$_{\sun}$/year.
The values of Eddington ratio range from 0.001 to $\sim$0.5 and peak at 0.1.
}
{} 

\authorrunning{Caccianiga et al.}
\titlerunning{black hole masses of type~1 AGN in the XBS}
   \maketitle

\section{introduction}

The nuclear activity of an active galactic nucleus (AGN) is powered by 
the accretion of matter into the gravitational well of the central 
supermassive black hole (SMBH). It has now become clear 
that the majority 
of galaxies host an SMBH and that they must have experienced an activity phase 
during their lifetime (see Merloni \& Heinz 2012 for a review). 
Much observational evidence, like the SMBH mass-bulge relations (e.g. 
Magorrian et al. 1998; G\"ultekin et al. 2009), strongly suggest that this 
activity phase must have played a critical role in galaxy evolution. 
For these reasons, a better understanding of the accretion mechanism 
represents a fundamental step not only in improving our knowledge of the AGN 
physics, but also for general comprehension of the galaxy formation 
and evolution. 

X-rays offer a direct probe of the accretion mechanism since they are 
produced in the very inner part of the nucleus 
through a (still poorly understood) mechanism that probably 
involves the electrons in a ''hot" corona and the UV photons produced 
within the accretion disk (e.g. Haardt \& Maraschi 1991, 1993), thus carrying
direct information on the physics very close to the SMBH. 
The highly penetrating capability of X-rays often makes them the only 
tool for gathering direct information on the nuclear activity when the disk 
emission, peaked in the UV part of the spectrum, is absorbed and unobservable.

While X-ray observations of single sources can shed light on the 
complexity 
of the emission at these energies, a statistical approach based on large
samples offers the unique opportunity of studying the link between hot corona
and the phenomenon of accretion on the central SMBH 
(e.g. see Young, Elvis \& Risaliti 2010; Vasudevan \& Fabian 2009; 
Grupe et al. 2010, Lusso et al. 2012 and references therein). 
To this end, statistically complete and 
well-defined samples of AGNs equipped with X-ray spectral data and 
with a reliable estimate of the accretion parameters (SMBH mass, the absolute 
accretion rate, the accretion rate normalized to the Eddington limit) are
required. 

The recent availability of statistical relations (see Vestergaard 2009 for a 
review) that 
allow the systematic computation of the black hole mass on large numbers of 
AGN has made it possible to estimate black hole 
masses for very large samples of AGNs (usually optically selected): for 
instance, the last 
release of the SDSS QSO catalogue contains a mass estimate for more than 
100,000 AGNs 
(Shen et al. 2011). In spite of these large numbers, the samples that contain
information on both black hole masses and X-ray spectra are significantly
smaller. In particular, if we restrict attention  
to the hard X-ray energies (above 2 
keV), where the primary X-ray emission is best observed and studied, the 
largest samples available for this kind of study contain a few hundred 
objects at most. The largest samples are often built using X-ray data 
from the XMM-{\it Newton} archive combined with optical data that come from 
SDSS (Risaliti, Young \& Elvis 2009; Vagnetti et al. 2010), from the literature 
(Bianchi et al. 2009) or from dedicated observations (Lusso et al. 
2012; Grupe et al. 2010). 
A major problem affecting many samples is that they are often 
just a collection of 
sources available in both an X-ray and an optical catalogue so they do not 
necessarily represent a statistically complete and representative sample of 
AGNs. 

To limit the possible biases deriving from this kind of selection, we present
here a new data set containing black hole masses and accretion rates (both
absolute and normalized to the Eddington limit) for a well-defined 
flux-limited sample of X-ray sources selected from XMM-{\it Newton}, the Bright
Serendipitous Survey (XBS\footnote{
The XBS is one of the research programmes conducted by
the XMM-Newton Survey Science Center (SSC, see http://
xmmssc-www.star.le.ac.uk), a consortium of 10 international institutions,
appointed by the European Space Agency (ESA) to help the
XMM-Newton Science Operations Centre (SOC) in developing the software
analysis system, to pipeline process all the XMM-Newton data, and
to exploit the XMM-Newton serendipitous detections. The Osservatorio
Astronomico di Brera is one of the Consortium Institutes.
}, Della Ceca et al. 2004; Caccianiga et al. 2008). 
The XBS is now almost completely identified 
($>$98\%) after ten years of dedicated spectroscopic observations, and
it contains, by definition, XMM-{\it Newton} data of medium/good quality (from 
100 to 10$^4$ net counts) that has
allowed systematic X-ray spectral analysis for all the selected AGN  
(Corral et al. 2011). For most of 
the type~1 AGN contained in this sample, the optical/UV spectral energy 
distribution has been studied and a reliable estimate (i.e. not based on a
bolometric correction) of the bolometric 
luminosity has already been published (Marchese et al. 2012). 
In this paper we present the estimate of the black hole masses, 
using the single-epoch method. In a companion paper we will use these values,
combined with the results of the  X-ray  analysis, to study the 
statistical relationship between X-ray properties and the accretion rate on 
the central SMBH (Fanali et al. in prep).  

The structure of the paper is the following. In Section~2 we briefly describe
the XBS sample while in Sects 3 and 4 we present the derivation of 
black hole masses and accretion rates, respectively. In Section~5 we discuss 
how the presence of the radiation pressure can change the derived quantities, 
and in Section~6 we summarize results and conclusions.

We assume a flat $\Lambda$CDM cosmology with H$_0$=65, $\Omega_{\Lambda}$=0.7 
and $\Omega_{M}$=0.3.

\section{The XBS sample of type~1 AGN}
The {\it XMM-Newton} Bright Serendipitous Survey (Della Ceca et 
al. 2004; Caccianiga et al. 2008) 
is a wide-angle (28 sq. deg), high Galactic latitude ($|b|>$20 deg) 
survey based on the {\it XMM-Newton} archival data. It is composed of two 
samples that are both flux-limited 
($\sim$7$\times$10$^{-14}$ erg s$^{-1}$ cm$^{-2}$) 
in two separate energy bands: the  0.5-4.5 keV band (the BSS sample) and the 
4.5-7.5 keV band (the ``hard'' HBSS sample). A total of 400 sources have been
selected, 389 belonging to the BSS sample and 67 to the HBSS sample 
(56 sources are in common). Selection criteria and the general properties of 
the 400 objects are discussed in Della Ceca et al. (2004). 

\begin{figure}
   \centering
   \includegraphics[width=7cm]{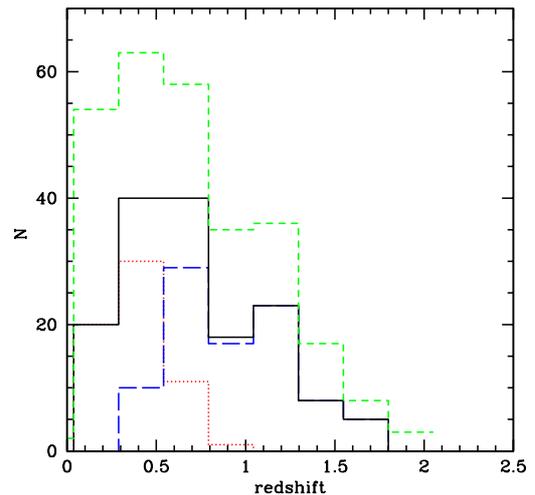}
\caption{Redshift distribution for the 154 XBS AGN1 discussed
in this paper  (continuous black line) compared to the distribution of the 
total sample of 276 AGN1 (green short-dashed line). 
Dotted (red) and long-dashed (blue) histograms indicate the objects whose 
black hole mass has been derived using the H$\beta$ and MgII$\lambda$2798\AA\ 
lines, respectively.}
         \label{hist_z}
   \end{figure}

To date, the spectroscopic identification  has  
nearly been completed, and 98\% of the 400 sources have been spectroscopicaly 
observed and classified. The details of the classification process are 
presented in Caccianiga et al. (2007, 2008). In this paper we want to derive
the mass of the central SMBH for the type~1 AGNs. In total, the XBS contains 
276 type~1 AGN but we have computed 
the M$_{BH}$ only for the sub-sample
of sources that was studied by Marchese et al. (2012) in order to have 
a reliable estimate of the bolometric luminosity. The sub-sample considered
by Marchese et al. contains the type~1 AGN that fall in the area of sky 
surveyed by GALEX (Martin et al. 2005; Morrissey et al. 2007),  
therefore it can be considered as representative 
of the entire XBS sample of type~1 AGN. We have then excluded a few sources
whose optical spectrum is either not available or without broad emission
lines required to compute the BH mass, leaving us with a total of 154 AGNs. 
In Fig~\ref{hist_z}, we compare the redshift distribution of the 154 type~1 
AGN studied here and of the total XBS sample of 276 type~1 AGN.  
The two distributions are similar, as demonstrated by a KS test 
(KS probability of 98.6\%).

\section{Black-hole mass}
To estimate the black hole masses of the XBS type 1 AGN, we used
the ``single epoch'' (SE) spectral method, which is based on measuring  
the broad line widths and the continuum emission in a single 
spectrum (e.g. see Peterson 2010 and 
Marziani \& Sulentic 2012 and references therein). 
The method assumes both that
the BLR traces the gravitational potential due to the presence of the 
central SMBH and that the virial theorem can be applied.
The two input quantities, the velocity dispersion and the size of the
system (R$_{BLR}$), can be inferred directly from the optical/UV spectrum: 
the line width yields direct
information on the velocity dispersion, while the continuum luminosity can
be used to estimate the system size through the R$_{BLR}$/L  
``scaling relations'' (e.g. Kaspi et al. 2000; Bentz et al. 2009). 
The unknown geometry of
the BLR is one fundamental source of uncertainty for this method and,
in general, for all methods based on the BLR kinematics (including
the reverberation mapping method, Vestergaard 2009). 
The average value of the ``virial factor'' 
that takes  the particular geometry of the system into account can be
assumed ``a priori'' (e.g. McLure \& Jarvis 2002) or it can be estimated 
through a comparison with the M$_{BH}$-$\sigma$ empirical relation 
observed in non-active galaxies (Onken et al. 2004, Woo et al. 2010, 
Graham et al. 2011). That the BLR geometry is probably different from 
object to object creates an
intrinsic dispersion on the ``virial factor'', which is one of the
most important sources of uncertainty associated to these methods. 
Besides this ``zero point'' uncertainty, the SE method has an additional
source of uncertainty due to the scatter on the size-luminosity relation.
All considered, the total uncertainty on the SE method has 
been recently estimated to be between 0.35 and 0.46 dex (Park et al. 2012).

The emission lines used for the M$_{BH}$ measurement depend on the 
redshift of the source. For the XBS sample, the type~1 AGNs
cover a redshift range between 0.02 and 2, therefore, the emission
lines that can be used for the mass estimate are the H$\beta$ 
(up to z$\sim$0.8) and the MgII$\lambda$2798\AA\ (from z$\sim$0.3). In a
number of cases both lines are included in the observed spectral range.

In this paper we adopt the relationships that are anchored to the virial
factor estimated by Onken et al. (2004).
For the H$\beta$, we used the relation discussed in 
Vestergaard \& Peterson (2006):

\begin{equation}
Log M_{BH} = 6.91 + 2 Log \frac{FWHM (H\beta)}{1000 km/s} + 0
.50 Log \frac{\lambda L_{5100\AA}}{10^{44} erg/s}
\end{equation}

For the MgII$\lambda$2798\AA\ line we used the relation presented in 
Shen et al (2011):

\begin{equation}
Log M_{BH} = 6.74 + 2 Log \frac{FWHM (MgII)}{1000 km/s} + 0.62 
Log \frac{\lambda L_{3000\AA}}{10^{44} erg/s}
\end{equation}

\noindent
this equation has been obtained by Shen et al. (2011) in such a way 
that the zero-order point (i.e. the virial factor) is the same as in the 
H$\beta$ relation presented above (eq. 1) so that the masses are
consistently derived from these two equations. 
In both relations, the line widths refer to the broad component, and it 
is assumed that a narrow component has been 
subtracted during the fitting procedure. 

In the following sections we describe in detail the methods adopted to compute
the two critical 
input quantities of the equations reported above, i.e. the line widths and the 
continuum luminosity. 

\subsection{Line width measurements}

The different dependence of M$_{BH}$ on line width and luminosity
(see eq. 1 and 2) means that the statistical 
(i.e. not including the intrinsic dispersion of the relation and the 
uncertainty on the virial factor) uncertainty
of the final M$_{BH}$ estimate will mostly come from the uncertainty on the
line width. The line width measurement is then particularly difficult
owing to the presence of different spectral components and considering the
average quality of our spectra (average S/N$\sim$10-11 in the spectral 
regions close to H$\beta$ and 
MgII$\lambda$2798\AA\ emission lines, with $\sim$25\% of objects having 
S/N below 5). 

In particular, the correct determination of the width of the broad component 
of the emission line is hampered by a narrow component
(which is particularly important for the H$\beta$ line) and by the 
iron pseudo-continuum (which is critical 
for the MgII$\lambda$2798\AA\ line). A simple component fit,  
not considering the possible presence of a narrow component, 
would lead to a systematic under-estimate of the broad line width (Denney
et al. 2009). 
At the same time, not considering the existence of the iron pseudo-continuum
may lead to an over-estimate of the line width.
A common practice for taking this spectral complexity  into account is to
subtract a FeII template  from the
spectrum and, then, fit the subtracted spectrum with a number of narrow and
broad components (usually with a Gaussian profile, e.g. see Shen et al. 2011
for details on the method).
In the following, we discuss separately the methods used to derive the width
of the broad components of the H$\beta$ and MgII.
                                           
\subsubsection{H$\beta$}
\begin{figure}
   \centering
   \includegraphics[width=7cm, angle=-90]{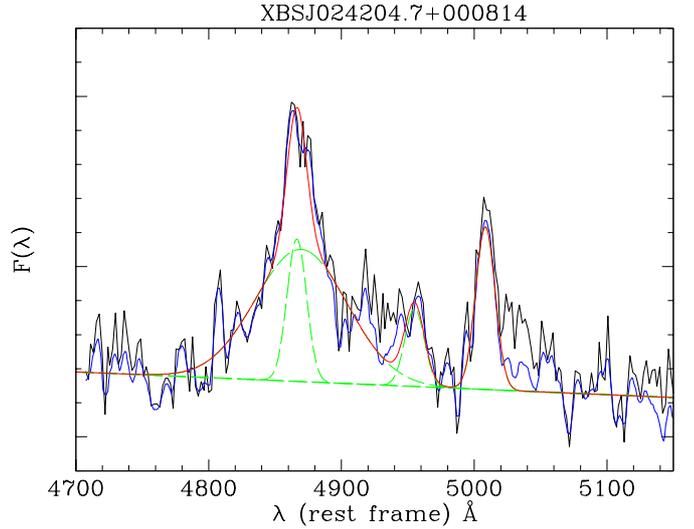}
      \caption{Example of a spectral model used to fit the 
region around the H$\beta$ line. As described in the text, we first
subtract an iron template from the spectrum (black line) and then 
we fit the residual (blue line) with a power-law continuum plus 3 Gaussians 
describing the narrow H$\beta$ and the two [OIII] lines, plus an
additional Gaussian to describe the broad 
component of the H$\beta$ line. These components are represented by the 
dashed green lines while the total fit is represented by the red continuous
line.}
        
         \label{ex_hb_fesubtracted}
   \end{figure}

For the fit of the H$\beta$ line we use the method usually adopted in the 
literature i.e. we subtract an iron template to the spectra and then fit 
the residuals. To this end, we use the iron template presented in 
V{\'e}ron-Cetty, Joly \& V{\'e}ron (2004) 
and consider the 3500-6000\AA\ (rest-frame) 
spectral region. In this 
procedure there are three independent parameters that need to be determined: 
the normalization of the iron template (N$_{Fe}$), the line broadening 
($\sigma_{Fe}$), and velocity offset (V$_{Fe}$) of the iron lines. 
Constraining the lattest two parameters is usually difficult even for 
good quality spectra. In medium quality spectra (like the one of the 
SDSS spectra considered by Shen et al. 2011, where S/N$\sim$10) these
parameters are poorly constrained (e.g. see discussion in Shen et al. 2011).
The quality of our spectra is, on average, similar to the ones of the SDSS
spectra (and in some cases even lower), so we decided to fix 
both parameters. 
After subtracting of the iron template, we fit the resulting
spectrum around the H$\beta$ line using a model composed by three components: 
a PL continuum plus four Gaussians
representing, respectively, the narrow and the broad components of the 
H$\beta$ and the two [OIII] narrow lines. 
The width of the component describing the narrow 
H$\beta$ is constrained to be equal to the width of two [OIII] lines. 
We run the fit in two steps: first we freeze
the positions of the Gaussians to the expected wavelengths. In a second step,
we leave the positions of the Gaussians describing the emission lines 
free to vary (with the maximum possible variation in the 
position of the iron components fixed to $\sim$30 \AA\ to avoid
problems with the fitting procedure).
The broad and the narrow H$\beta$ components 
do not necessarily peak at the same wavelength to account for possible 
velocity offsets between the BLR and the NLR. 
We show an example of this fitting procedure in Fig.~\ref{ex_hb_fesubtracted}
%

\begin{figure}
   \centering
   \includegraphics[width=7cm, angle=0]{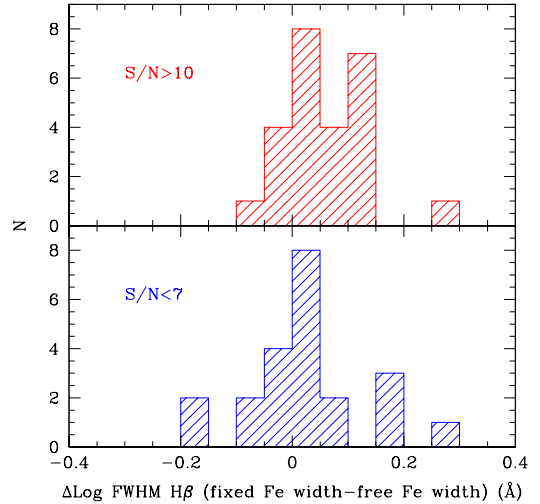}
      \caption{Difference between the logarithms of the broad H$\beta$ 
emission line width derived with two different methods, one based on the
subtraction of an iron template where the line widths are fixed and 
a second method that, instead, leaves the iron line widths free to vary
(see text for details). The data are split on the basis of the 
signal-to-noise around the H$\beta$ line.}        
         \label{SN}
   \end{figure}

We note that keeping the iron line width and position
fixed during the fitting procedure may introduce a certain degeree of
uncertainity (even systematic) in the computation of the broad H$\beta$
width. The possibity that the iron lines could be systematically shifted with
respect to the H$\beta$ line has been investigated by several authors, 
and although there have been some claims of systematic large velocity offsets 
(up to 2000 km s$^{-1}$) in the spectra of SDSS quasars (Hu et al. 2008), 
the analysis of good S/N spectra has recently demonstrated that the 
actual offsets are much smaller ($<$300 km $^{-1}$, Sulentic et 
al. 2012), if present. The impact
of such small offsets on our fitting procedure is not going to be relevant. 
On the contrary, keeping the iron line width fixed can have a 
more significant impact on our mass estimates.  
To  quantitatively evaluate this effect,   
we  applied a second fitting 
method, not based on subtraction of an iron template, using an approach
similar to the one used for fitting the MgII$\lambda$2798\AA\ line 
(see below). 
In this method, we adopted a model composed of six Gaussians plus a power-law  
continuum. Two Gaussians are used to model the H$\beta$ (for the narrow and 
the broad components), while two Gaussians are used to fit the 
two [OIII] narrow emission lines. The remaining two Gaussians are used to 
account for the two strongest FeII components usually observed at 4924\AA\ and 
5023\AA. The widths of these two lines are left free to vary.
Then we ran the fitting procedure following the two steps described 
above and found the best-fit width of the broad H$\beta$ component. 
The resulting widths were finally compared to those obtained by 
subtracting the iron template. 
We carried out this comparison by splitting the sample into  
two sets: a first data set containing
only the low S/N (in the H$\beta$ region) spectra (S/N$<$7) and a second data
set containing the best spectra we have (S/N$>$10). In Fig.~\ref{SN}
we show the distribution of the difference in the widths (in Log) 
computed using the two procedures, for both data sets. In the
case of low S/N spectra there is no systematic difference between the two
estimates. This is expected since, in case of very poor quality spectra, it is
very difficult to detect any real  difference in the iron line width, and all
the differences are probably due just to random fluctuations in the fitting 
procedure. In contrast, for relatively good spectra (S/N$>$10), 
we do observe a significant ($\sim$3$\sigma$) systematic 
offset between the line widths,
the H$\beta$ being typically larger in the iron template subtraction method 
(where the iron lines are fixed), when compared to the method where the iron
lines are left free to vary. This is probably because, in the 
first method, part of the iron emission may be included in the broad 
H$\beta$ component thus producing larger widths. We stress, however, that
even in the S/N$>$10 data set, the average quality of the spectra (S/N 
between 10 and 30) is certainly
not comparable to the one typically required for a proper 
spectral deconvolution ($>$50) and, therefore, there is a high 
degree of degeneracy in the fitting process. We cannot exclude, for instance, 
that part of the observed offset is related to an underestimate of the 
broad H$\beta$ component in the method where the iron widths are left free 
to vary. 
For this reason, it is difficult to establish which one of the two methods
gives better results. However, the observed offset can be used as an 
estimate of the possible effect on the broad H$\beta$ width because  
we have fixed the 
iron width when subtracting the iron template. The observed offset is
0.057 dex, which translates into an expected offset in the mass computed using
the H$\beta$ line of $\sim$0.11. 
This offset is within the average statistical uncertainty on the 
masses computed from the H$\beta$ line ($\sim$0.18 dex).

%

We finally note that fitting the H$\beta$ broad line using only one Gaussian 
is 
certainly a simplification. The analysis of high S/N spectra of local
Seyfert galaxies has revealed a complex phenomenology (e.g. see Sulentic, 
Marziani \& 
Dultzin-Hacyan 2000 for a review).
Given the
typical S/N of our spectra, however, any attempt to provide a more 
complex fit
to the broad H$\beta$ profile would lead to very uncertain results, except
for very few cases. Indeed, this is a general problem connected with the
systematic application of the SE relation to large samples of spectra
whose quality is typically much lower than that of the brightest and 
best-studied local Seyferts. 

\subsubsection{MgII}
\begin{figure}
   \centering  
   \includegraphics[width=7cm, angle=-90]{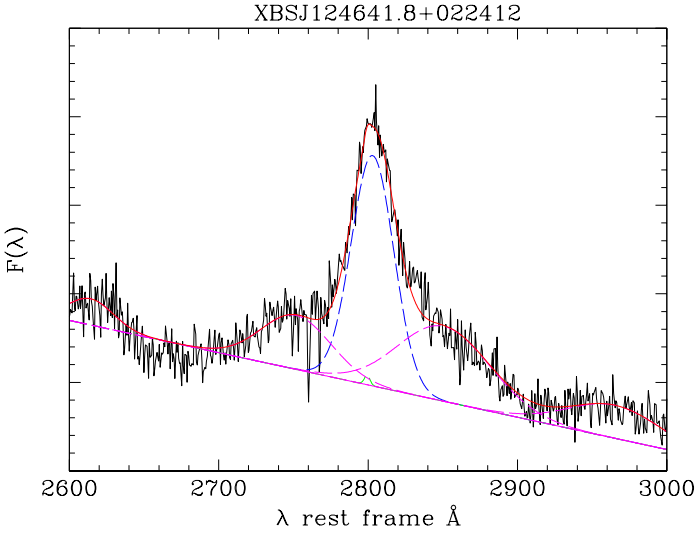}
      \caption{Example of a spectral model to fit the 
region around the  MgII$\lambda$2798\AA\ line. 
This method includes the iron lines directly in the fitting
procedure rather than subtracting an iron template from the spectrum, as 
typically done in the literature. 
The total fit is represented by the solid red line while the
different components (the power-law continuum, the narrow and the broad 
components of the line and the iron humps) are represented by the 
dashed lines.}
        
         \label{ex_mg_fe}
   \end{figure}

For the MgII we did not follow the same procedure as adopted for the
H$\beta$ line due to the
difficulty of obtaining a reliable iron template at these wavelengths. 
We thus decided to 
include the iron components in the fitting procedure. 
Specifically, we adopted a model 
including two Gaussians for the narrow and broad components of 
MgII$\lambda$2798\AA\ plus four additional Gaussians to reproduce the iron 
humps 
at 2630\AA, 2740\AA, 
2886\AA\, and 2950\AA\, plus a power-law continuum (see Fig.~\ref{ex_mg_fe}).

%

Since, in the case of MgII, we do not have the
two [OIII] line as a reference for the narrow line widths, we set the
MgII narrow component to be equal to the instrumental resolution, for 
the spectra with a resolution worse 
than 500 km/s. For the very few spectra
with better resolution, the width of the narrow component is fixed to 
500 km/s. Again, as a first step we fix the positions
of the components to the expected values and, then, we left them free to vary
(with a maximum possible variation of 30\AA\ for the iron components).
In fitting the MgII line we have thus assumed that a 
narrow component is present. It should be noted, however, that  
for the MgII$\lambda$2798\AA\ line, the actual presence of a narrow
component is less obvious than for the H$\beta$ line. 
In their work, Vestergaard \& Osmer (2009) did not 
subtract a narrow component during the fitting procedure of the MgII 
profile (which was modelled with two Gaussians both attributed to the broad 
component), while other authors (e.g. Mc Lure \& Dunlop 2004) have considered 
a narrow plus a broad component for the MgII$\lambda$298\AA\ line as in the 
analysis presented here. The choice of including the narrow component
of the MgII$\lambda$2798\AA\ is somewhat arbitrary. In our analysis,  
including the narrow MgII$\lambda$2798\AA\ component gives a slightly 
better consistency between the masses computed using 
MgII$\lambda$2798\AA\ and those computed using H$\beta$, so we 
decided to adopt this type of model. 

\subsubsection{Instrumental resolution}
Finally, given the moderate resolution of the spectroscopic observations 
($\sim$650-1200 km s$^{-1}$), we applied a correction to the widths of
the broad components of both H$\beta$ and MgII$\lambda$2798\AA, 
resulting from the fitting procedures described above, to account for the 
instrumental broadening, i.e.,

\begin{center}
$\Delta\lambda = \sqrt{\Delta\lambda_o^2 - \Delta\lambda_{inst}^2}$
\end{center}

\noindent
where $\Delta\lambda$, $\Delta\lambda_o$ and $\Delta\lambda_{inst}$ are the 
intrinsic, the observed, and the instrumental line width,  
respectively. 

\subsection{Monochromatic luminosities}
Determination of the monochromatic luminosities at 5100\AA\ and 
3000\AA\ also requires some caution. In principle we can use the fluxes 
derived directly from the spectra. This procedure, however, is not
accurate for several reasons:

\begin{itemize}

\item the absolute spectro-photometric calibration of our spectra is not 
always accurate since most of the data have been collected during
non-photometric nights;

\item  the spectra are often contaminated by the host galaxy light (the 
slit width used was often relatively large, from 1 to 2 arcsecs, 
depending to the seeing conditions);

\item the spectra must be corrected for the extinction, both Galactic and at
the source. This is a particularly critical point since, given the
relatively hard X-ray selection band, the XBS sample contains many type~1 
AGNs with moderate levels of absorption  (A$_V$ up to 1-2 magnitudes, see
Caccianiga et al. 2008).

\end{itemize}

To account for these points, we used the result of a 
systematic study of the optical/UV  spectral energy distribution (SED) 
of the type~1 AGN of the XBS survey, described in Marchese et al. (2012). 
In this work we have collected photometric points, both in the optical 
(most from the SDSS) and in the UV band (from GALEX) and built the SED for
each source. 
In the derivation of the SED we carefully took the
presence of the host galaxy into account, on the basis of the strength of the
4000\AA\ contrast, and excluded it from the final SEDs. We also
corrected the photometric points for the extinction, both due to our 
Galaxy and at the source, 
using the values of N$_H$ derived from the X-ray
analysis (Corral et al. 2011) and assuming a Galactic gas-to-dust ratio. 
This is certainly an approximation since there are well-known examples 
of AGN where the dust-to-gas ratio is significantly different from what
is observed in our Galaxy. However, in the XBS survey we have found generally
good agreement between the optical classification (type1/type2 AGN) and the
measured levels of N$_H$ (lower or greater than 4$\times$10$^{21}$ cm$^{-2}$), 
with only a few ($<$10\%) exceptions (Caccianiga et al. 2004; Corral et al. 
2011). Therefore, we expect that this problem is not
going to have a strong impact on our results, at least from a statistical
point of view.

These SEDs have been then fitted with a multi-colour
black-body accretion disk model, which includes corrections for temperature
distribution near the black hole (for details see DISKPN in the XSPEC 12 
software package, Arnaud et al., 1996).  
From this fit, we computed the rest frame 5100\AA\ and 3000\AA\ 
luminosities to be used in eq.~1 and 2 for the mass estimate. 

\subsection{Computing the BH masses}
Using the methods described in the previous sections, we computed the
black hole masses for all the 154 type~1 AGNs of the XBS for which we
analysed 
the SED, as described in Marchese et al. (2012) and for which we acquired 
an optical
spectrum. For 32 objects we only covered the H$\beta$ emission line while 
for 70 objects we have covered only the MgII$\lambda$2798\AA\ line. In 52 
cases 
we have detected both lines in the spectrum. In these cases we chose the
mass estimate that is considered more accurate, i.e. the one based on the 
line with the best S/N and/or with the smallest error in the measured 
width (quite
often, one of the two lines is at the edge of the observed spectrum). 
Overall, the black hole masses were derived from the H$\beta$, in 62 
cases, and from MgII$\lambda$2798\AA\ line, in 92 cases.  

\begin{figure}
   \centering
   \includegraphics[width=7cm]{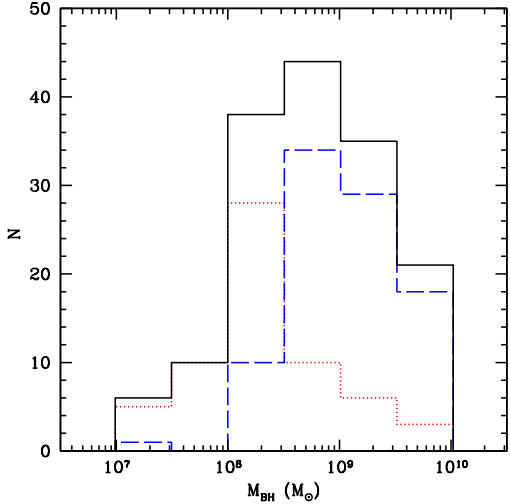}
\caption{Distribution of the black hole masses for the 154 XBS AGN1.
Dotted (red) and dashed (blue) histograms show masses derived from
H$\beta$ and MgII$\lambda$2798\AA\ lines respectively.}
         \label{hist_mass}
   \end{figure}

The masses for the 154 type~1 AGN are reported in 
Table~\ref{table}, 
together with the (statistical) errors. In Table~\ref{table} we also 
report 
the FWHM of the lines and the values of the monochromatic luminosities used 
for the mass estimate.
The distribution of the masses obtained for the 154 AGN1 of the XBS sample are 
reported in Fig.~\ref{hist_mass}.

\subsection{Uncertainties on BH Masses}
Statistical uncertainities on the BH masses were estimated by combining 
the statistical errors on both line width and monochromatic luminosity, 
assuming that the two errors are independent:

\begin{center}
$\sigma^{+,-}_{LogM}$ = $\sqrt{(A\sigma^{+,-}_{Log FWHM})^2 + (B\sigma^{+,-}_{LogL})^2}$
\end{center}

\noindent
where A=2 and B is equal to 0.5 for the H$\beta$ while it is 0.62 for 
MgII$\lambda$2798\AA. 
$\sigma^{+,-}$ are the asymmetric errors (at the 68\% confidence level) to 
the logarithm of the FWHM and luminosities, respectively.

The errors on the H$\beta$ and MgII$\lambda$2798\AA\ broad components are 
derived from the fitting precedure described above, by imposing 
$\Delta \chi^2$=$\pm$1.
Similarly, the errors on the monochromatic luminosities are computed
from the SED fitting procedure by again imposing $\Delta \chi^2$=$\pm$1 from
the best-fit value. As described in Marchese et al. (2012), the SED fitting 
procedure takes the errors on the photometric points into account. These 
1$\sigma$ uncertainities include both the errors on photometry and additional 
sources of error due to the correction for the intrinsic extinction and 
the long term variability (since the used photometric data are not 
simultaneous). The uncertainty due to the correction for the host galaxy,
based on the 4000\AA\ break, is not folded into these errors. However,  
in Marchese et al. (2012) we evaluated that by changing the starting 
value of the 4000\AA\ break within a reasonable range of values 
(from 45\% to 55\%), the variations in the photometric points 
only produce a negligible ($\leq$14\%) change in the best fit luminosity. 

The statistical 1-$\sigma$ errors on the broad line widths, monochromatic 
luminosities, and on the final black hole masses are reported in 
Table~\ref{table}.
We stress that the errors on black hole masses 
do not include the uncertainity on the
SE method that, as already explained, is expected to be between 
0.35 dex and 0.46 dex (Park et al. 2012) i.e. dominant when compared to 
the average statistical errors ($\sim$0.14 dex).

\subsection{Comparison of the black hole mass estimates}
With the derived line widths and monochromatic luminosities we computed 
the M$_{BH}$ for all the AGN1 for which either the H$\beta$ or the
MgII$\lambda$2798\AA\ lines have been observed. 
   \begin{figure}
   \centering
   \includegraphics[width=7cm]{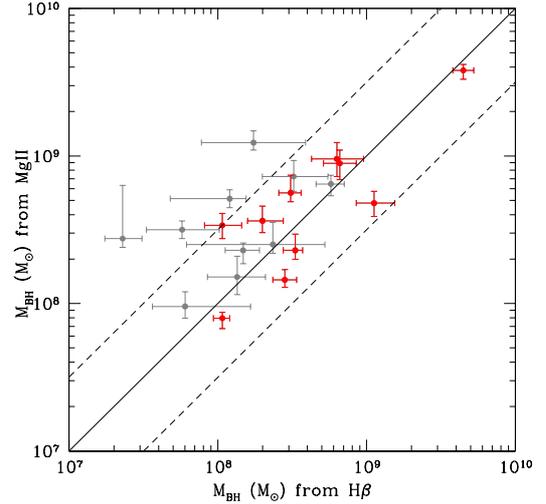}
 
\caption{Comparison between black hole masses computed on the basis of the 
MgII$\lambda$2798\AA\ and the H$\beta$ lines for the XBS AGNs where both
lines are included in the spectrum. Red points represent sources with a 
relatively high S/N ($>$5) around the line of interest and with 
lower statistical error bars ($<$0.2 dex) while grey points are objects with
lower S/N spectra and/or larger error bars. 
As reference we plot the relation 1:1 
(solid line), while the two dashed lines represent a scatter of 0.5 dex.}
         \label{comparison_hbmg}
   \end{figure}
For the 52 sources where both H$\beta$ and MgII$\lambda$2798\AA\ are 
included in the spectrum it is possible to compare the two
M$_{BH}$ estimates. To evaluate the presence systematic
offsets better, we first considered the 
objects with a relatively good spectrum (S/N$>$5) and excluded the sources
with large statistical errors on the final mass estimate ($>$0.2 dex). 
The comparison (Fig.~\ref{comparison_hbmg}) 
shows generally good agreement, 
without significant offsets and with a spread of $\sim$0.28 dex. Considering
all the objects, including those with less accurate determination of the
mass the spread increases to $\sim$0.38 dex, and there seems
to be a systematic shift probably related to the difficulty of properly 
accounting for all the components during the spectral fit (in particular 
the iron lines around the MgII$\lambda$2798\AA\ line and the narrow component
of the H$\beta$ line). In Table~\ref{table} we have flagged the masses 
derived from a problematic fit and those resulting from the analysis of poor 
S/N ($<$5) spectra.

\begin{figure}
   \centering
   \includegraphics[width=7cm]{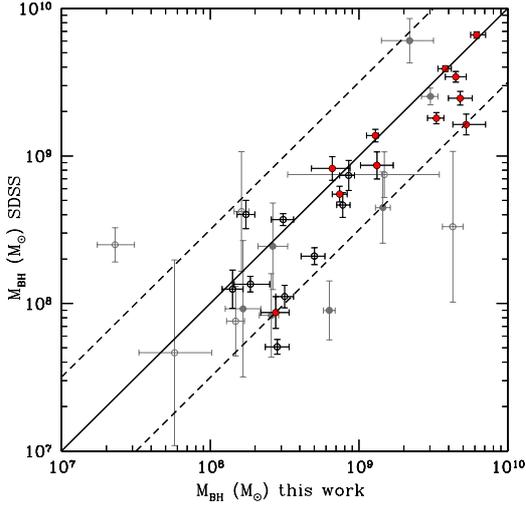}
\caption{Comparison between black hole masses computed in this 
paper and those computed by Shen et al. (2011), for the XBS AGNs 
included in the SDSS sample. Red and black points represent the objects
with higher signal-to-noise ratio ($>$5) and smaller uncertainties on the
mass derivation ($<$0.2 dex). 
Grey points, instead, represent the data with lower S/N
and/or larger error bars. 
Filled (and red, in electronic version) points are sources for which we have 
used the  SDSS spectrum to derive 
the BH mass, while open points indicate sources for which we used
an optical spectrum taken in our own observations.  
As reference we plot the relation 1:1 
(solid line), while the two dashed lines represent a scatter of 0.5 dex}
         \label{conf_sdss}
   \end{figure}

As a further test of the reliability of our mass estimate 
we compared the black hole masses derived in our work with those 
computed in Shen et al. (2011) for the few sources in common. Since Shen
et al. (2011) presents masses computed using different formulae, 
we used  the ones computed in the same way for the comparison,
i.e. the VP06 for H$\beta$, and the S10 for MgII$\lambda$2798\AA.
The result of the 
comparison is presented in Fig.~\ref{conf_sdss}. In some cases, we used 
the same SDSS spectrum to derive the BH masses while in other 
cases we acquired an independent spectrum.
As before, we first excluded from the test the sources with low
S/N ($<5$) spectra (used in our analysis) and large errors 
($>$0.2 dex) in either our estimate or in the Shen et al. estimate. 
The comparison shows a spread of $\sim$0.2-0.3 dex and a marginal 
systematic offset 
between the two masses, with the ones computed in this work being larger on 
average by a 
factor $\sim$0.17 dex. The offset is mainly present in the masses computed 
from 
MgII$\lambda$2798\AA.  
By comparing separately the line widths and the monochromatic luminosities 
we
have established that this offset is mainly attributed to an offset in 
luminosity
rather than in line width. This offset is probably due to the method we 
used to compute the
monochromatic luminosities that corrects for the extinction (both Galactic 
and at the source), 
as explained in the previous sections, thus yelding, on average, to higher  
corrected luminosities,
in particular in the blue/UV spectral region. 
Considering all the sources in common between the two samples the spread 
increases to $\sim$0.4 dex.

%

\section{Eddington ratio and $\dot{M}$}

An important parameter that is suspected of regulating a number of 
observational
properties of AGNs is the ``normalized'' bolometric luminosity, i.e. the 
so-called Eddington ratio, which is defined as

\begin{equation}
\lambda=L_{bol}/L_{Edd}
\end{equation}

\noindent
where:

\begin{equation}
L_{Edd}=1.26\times10^{38} \frac{M_{BH}}{M_{\sun}} erg s^{-1}
\end{equation}

\begin{figure}
   \centering
   \includegraphics[width=7cm]{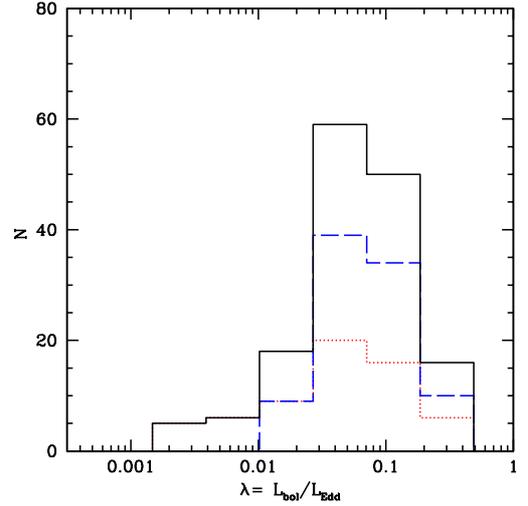}
\caption{Distribution of the values of Eddington ratio for the 154 XBS AGN1.
Dotted (red) and dashed (blue) histograms show the values based on masses 
derived from H$\beta$ and MgII$\lambda$2798\AA\ lines respectively.}
         \label{hist_eddrat}
   \end{figure}

We compute the values of Eddington ratio using the bolometric luminosities 
taken from Marchese et al. (2012) which was computed, as 
explained above, by fitting the optical/UV data
with a disk model. The photometric points, and therefore the bolometric 
luminosity, were corrected for reddening as detailed in Marchese et al
(2012). The distribution of Eddington ratios is 
reported in Fig.~\ref{hist_eddrat}

From the bolometric luminosity we can also derive an estimate of the
absolute (i.e. not normalized to the Eddington limit) accretion rate:

\begin{equation}
\dot{M} = \frac{L_{bol}}{\eta c^2} \sim1.8\times 10^{-3}\frac{L_{44}}{\eta} M_{\sun}  yr^{-1}
\end{equation}

\noindent
where L$_{44}$ is the bolometric luminosity in units of 10$^{44}$ erg s$^{-1}$ and $\eta$
is the efficiency of the mass-to-energy conversion. We assume here an 
efficiency of 0.1 (Marconi et al. 2004).
We note that the bolometric luminosities used to compute $\dot{M}$ also 
include the X-ray emission (in addition to the disk component)
as described in Marchese et al. (2012). Therefore, by using these bolometric
luminosities to compute $\dot{M}$ we are implicitly 
assuming that the energy budget carried by the X-ray emission is directly 
related to the accretion process. This is, of course, 
not an obvious assumption, 
since the origin of the X-ray emission is still an open issue. 
In any case, we stress that the contribution of the X-ray emission to the 
bolometric luminosity is, in general, relatively low ($\sim$25\% on 
average in our sample) and, therefore, the values of $\dot{M}$ are 
not going to change significantly (on average) 
if we use only the disk emission in eq. (5).

   \begin{figure}
   \centering
   \includegraphics[width=7cm]{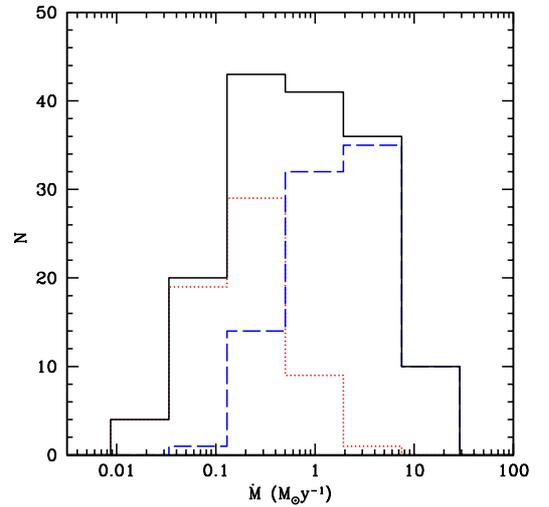}
\caption{Distribution of the values of $\dot{M}$ for the 154 XBS AGN1.
Line styles as in Fig.~\ref{hist_mass}}
         \label{hist_mdot}
   \end{figure}

The distribution of $\dot{M}$ is reported in Fig.~\ref{hist_mdot}. To 
facilitate the comparison with previous figures we also show the 
$\dot{M}$ separately for H$\beta$ and MgII$\lambda$2798\AA\
mass-derived sources, although in this case, the value of $\dot{M}$ does not 
depend on the derived BH mass.

\section{The effect of radiation pressure}
It has been suggested (Marconi et al. 2008; Marconi et al. 2009) 
that the black hole masses derived from the virial theorem can be severely
underestimated due to the effect of the radiation pressure. This effect,
not considered in the usual SE relations, is expected to be important for
accretion rates close to the Eddington limit according to the following
equation (Marconi et al. 2008):

\begin{equation}
M_{BH}=M_{BH,0}[1+\lambda_0(1-a+\frac{a}{\sigma_T N_H})]
\end{equation}

\noindent
where M$_{BH}$ is the ``real'' black hole mass, M$_{BH,0}$ is the black hole
virial mass computed by neglecting the radiation pressure, 
$\lambda_0$ is the Eddington ratio computed using M$_{BH,0}$, 
$a=L_{ion}/L$ (i.e. the ratio between the ionizing continuum 
luminosity and the bolometric luminosity), $\sigma_T$ is the Thomson 
cross-section, and N$_H$ the column density of each BLR cloud along the
line of sight. As noted by Marconi et al. (2008), for reasonable assumptions on
the BLR density ($\sim$10$^{23}$ cm$^{-2}$) if the accretion is close to 
the Eddington
limit, the correction could be as high as a factor 10. 
The actual importance of the radiation pressure, however, has been 
debated in the recent literature. Netzer (2009), for instance, notes that
the Eddington ratios of a sample of type~1 AGN from the SDSS (whose 
black hole masses were computed using the virial method), when corrected
for the radiation pressure, turnes out to be significantly lower when compared 
to the Eddington ratio distribution of an SDSS sample of type~2 AGN 
for which the black hole masses have been computed using a different 
technique (M-$\sigma$ relation). In contrast, if no correction is 
applied, the two distributions are similar.

Given the difficulty of assessing the actual importance of the radiation 
pressure, we decided to present  both the corrected and the
uncorrected masses and Eddington ratios in this paper. 
The corrected masses, in particular,
were computed using the equation above and assuming a=0.6 and 
N$_H$=10$^{23}$ cm$^{-2}$ (the values assumed in 
Marconi et al. 2008).

\begin{figure}
   \centering
   \includegraphics[width=7cm]{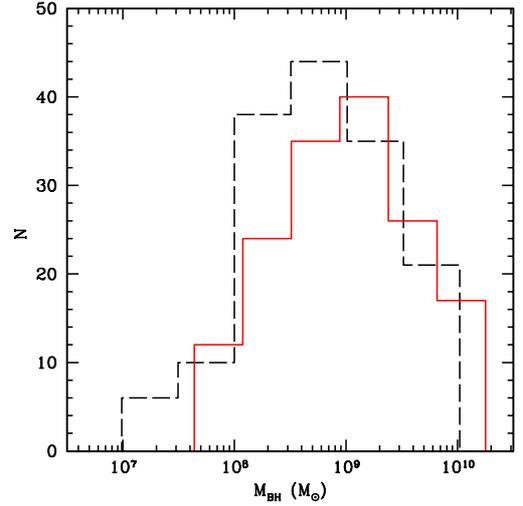}
\caption{Distribution of the black hole masses corrected for the radiation
pressure (red continuous line), as described in the text. For comparison
we show the distribution of uncorrected masses (black dashed line)}
         \label{marconi_bh}
   \end{figure}

\begin{figure}
   \centering
   \includegraphics[width=7cm]{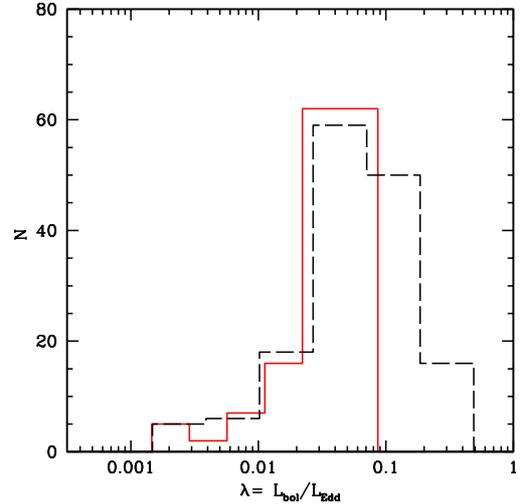}
\caption{Distribution of the Eddington ratios corrected for the radiation
pressure (red continuous line), as described in the text. For comparison
we show the distribution of uncorrected masses (black dashed line)}
         \label{marconi_edd}
   \end{figure}

In Fig.~\ref{marconi_bh} we show the black hole mass and in 
Fig.~\ref{marconi_edd} the Eddington ratio distributions corrected for the
radiation pressure and compared with the uncorrected quantities.
As expected, the corrected masses are shifted towards the higher values, 
while the Eddington ratio presents a sharp cut off at ~0.1 (see discussion
in Marconi et al. 2008). 

The values of masses and Eddington ratios corrected for the radiation
pressure are included in Table~\ref{table}.

\section{Summary and conclusions}

We have presented black hole masses and accretion rates (both absolute and 
relative to the Eddington limit) for 154 type~1 AGNs belonging to the 
XBS sample. The masses were derived using the single-epoch method and
adopting the most recent scaling relations involving the H$\beta$ and the 
MgII$\lambda$2798\AA\ emission lines.
The selected sources cover a range of masses from 10$^{7}$ to 10$^{10}$ 
M$_{\sun}$ with a peak around 8$\times$10$^{8}$ M$_{\sun}$ and a range of 
accretion rates from 0.01 to $\sim$50 M$_{\sun}$/year (assuming an efficiency 
of 0.1), with a peak at around 1 M$_{\sun}$/year.
The values of the Eddington ratio range from 0.001 to $\sim$0.5 and peak at 
0.1.

We have verified that the computed masses are in broad agreement with the ones 
presented in Shen et al. (2011) although we found a systematic 
offset of $\sim$0.17 dex (with our masses being higher) 
probably because of the different methods adopted in the two works 
to estimate the continuum luminosity.

We stress that the 154 type~1 AGN presented here constitute a well-defined 
flux-limited sample of type~1 AGN and not just a collection of data from the 
literature or from public archives. This characteristic, combined with the
systematic availability for all these objects of crucial X-ray information 
(based on X-ray spectral analysis) and on the optical/UV SED, makes this
sample instrumental for statistical studies. 
In a companion paper (Fanali et al. in prep), 
we will use the results presented here to study 
the link between the hot-corona, responsible for the X-ray 
emission, and the accretion process onto the central black hole. 

\section*{Acknowledgments}
We thank the referee for useful comments that significantly improved the paper 
and Tommaso Maccacaro for his initial involvement in the XBS project.
The authors acknowledge financial support from ASI (grants n.~I/088/06/0
and I/009/10/0).
SM acknowledges financial support by the Spanish Ministry of Economy and 
Competitiveness, through grant AYA2010-21490-C02-01.

{}

\begin{landscape}
\begin{table}
\caption{black hole masses of the XBS AGN1}
\label{table}
\begin{tabular}{r r r r r r r r r r r r r r}
\hline\hline
name & z & LogFWHM & LogFWHM  & log$\lambda$L$_{\lambda}$ & Log$\lambda$L$_{\lambda}$ & LogM$_{BH}$ & LogM$_{BH}$ & logM$_{BH}$ & logM$_{BH}$ & Log$\dot{M}$ & LogL/L$_{Edd}$ & LogL/L$_{Edd}$& Flag\\
     &   & H$\beta$ & MgII$\lambda$2798\AA & 5100\AA & 3000\AA& H$\beta$&MgII$\lambda$2798\AA  & best & p$_{rad}$  & & & p$_{rad}$ & \\
(1) & (2) & (3) & (4) & (5) & (6) & (7) & (8) & (9) & (10) & (11) & (12) & (13) & (14) \\
\hline
XBSJ000027.7$-$250442 & 0.336 & 3.81$^{+0.045}_{-0.043}$ & 3.72$^{+0.018}_{-0.019}$ & 
44.21$^{+0.11}_{-0.16}$ & 44.34$^{+0.11}_{-0.16}$ &
8.63$^{+0.10}_{-0.12}$ & 8.39$^{+0.07}_{-0.09}$ & 
8.63$^{+0.10}_{-0.12}$ &  8.68 & -0.94$^{+0.09}_{-0.12}$ & -1.93$^{+0.13}_{-0.17}$ & -1.98 &     \\
XBSJ000031.7$-$245502 & 0.284 & 3.48$^{+0.664}_{-0.042}$ & --$^{ }_{ }$ & 
44.28$^{+0.12}_{-0.48}$ & 44.17$^{+0.13}_{-0.48}$ &
8.02$^{+1.32}_{-0.25}$ & --$^{ }_{ }$ & 
8.02$^{+1.32}_{-0.25}$ &  8.15 & -1.05$^{+0.11}_{-0.33}$ & -1.43$^{+1.32}_{-0.41}$ & -1.56 &   2 \\
XBSJ000102.4$-$245850 & 0.433 & 3.73$^{+0.067}_{-0.062}$ & 3.66$^{+0.033}_{-0.017}$ & 
43.57$^{+0.14}_{-0.16}$ & 43.79$^{+0.15}_{-0.15}$ &
8.16$^{+0.15}_{-0.14}$ & 7.93$^{+0.09}_{-0.09}$ & 
8.16$^{+0.15}_{-0.14}$ &  8.26 & -1.06$^{+0.07}_{-0.06}$ & -1.58$^{+0.17}_{-0.15}$ & -1.68 &     \\
XBSJ001831.6+162925 & 0.553 & --$^{ }_{ }$ & 3.50$^{+0.007}_{-0.007}$ & 
45.09$^{+0.12}_{-0.09}$ & 45.29$^{+0.12}_{-0.09}$ &
--$^{ }_{ }$ & 8.54$^{+0.06}_{-0.05}$ & 
8.54$^{+0.06}_{-0.05}$ &  8.92 & 0.06$^{+0.10}_{-0.09}$ & -0.84$^{+0.12}_{-0.10}$ & -1.22 &   2 \\
XBSJ002618.5+105019 & 0.473 & 3.75$^{+0.045}_{-0.064}$ & 3.48$^{+0.036}_{-0.040}$ & 
45.23$^{+0.11}_{-0.10}$ & 45.43$^{+0.11}_{-0.10}$ &
9.03$^{+0.10}_{-0.14}$ & 8.59$^{+0.09}_{-0.09}$ & 
9.03$^{+0.10}_{-0.14}$ &  9.24 & 0.20$^{+0.10}_{-0.08}$ & -1.19$^{+0.14}_{-0.16}$ & -1.40 &     \\
XBSJ002637.4+165953 & 0.554 & 3.42$^{+0.050}_{-0.238}$ & 3.72$^{+0.019}_{-0.020}$ & 
44.92$^{+0.12}_{-0.15}$ & 45.06$^{+0.11}_{-0.16}$ &
8.21$^{+0.11}_{-0.41}$ & 8.83$^{+0.07}_{-0.09}$ & 
8.21$^{+0.11}_{-0.41}$ &  8.63 & -0.20$^{+0.08}_{-0.11}$ & -0.77$^{+0.14}_{-0.42}$ & -1.19 &   2 \\
XBSJ003315.5$-$120700 & 1.206 & --$^{ }_{ }$ & 3.94$^{+0.030}_{-0.033}$ & 
45.32$^{+0.28}_{-0.34}$ & 45.52$^{+0.28}_{-0.34}$ &
--$^{ }_{ }$ & 9.56$^{+0.16}_{-0.18}$ & 
9.56$^{+0.16}_{-0.18}$ &  9.67 & 0.40$^{+0.21}_{-0.19}$ & -1.52$^{+0.26}_{-0.26}$ & -1.63 &     \\
XBSJ003316.0$-$120456 & 0.660 & --$^{ }_{ }$ & 3.62$^{+0.031}_{-0.033}$ & 
45.01$^{+0.18}_{-0.15}$ & 45.25$^{+0.18}_{-0.15}$ &
--$^{ }_{ }$ & 8.76$^{+0.11}_{-0.10}$ & 
8.76$^{+0.11}_{-0.10}$ &  9.09 & 0.19$^{+0.16}_{-0.14}$ & -0.93$^{+0.19}_{-0.17}$ & -1.26 &     \\
XBSJ003418.9$-$115940 & 0.850 & --$^{ }_{ }$ & 3.76$^{+0.034}_{-0.037}$ & 
44.70$^{+0.17}_{-0.23}$ & 44.94$^{+0.18}_{-0.22}$ &
--$^{ }_{ }$ & 8.84$^{+0.11}_{-0.13}$ & 
8.84$^{+0.11}_{-0.13}$ &  9.03 & -0.05$^{+0.14}_{-0.16}$ & -1.25$^{+0.18}_{-0.21}$ & -1.44 &     \\
XBSJ005009.9$-$515934 & 0.610 & 3.66$^{+0.172}_{-0.299}$ & 3.67$^{+0.075}_{-0.172}$ & 
44.44$^{+0.11}_{-0.10}$ & 44.63$^{+0.12}_{-0.09}$ &
8.45$^{+0.35}_{-0.58}$ & 8.47$^{+0.16}_{-0.34}$ & 
8.45$^{+0.35}_{-0.58}$ &  8.62 & -0.48$^{+0.08}_{-0.06}$ & -1.29$^{+0.36}_{-0.58}$ & -1.46 &   2 \\
XBSJ005031.1$-$520012 & 0.463 & 3.60$^{+0.110}_{-0.097}$ & 3.75$^{+0.028}_{-0.030}$ & 
44.85$^{+0.19}_{-0.26}$ & 45.05$^{+0.19}_{-0.27}$ &
8.53$^{+0.23}_{-0.22}$ & 8.89$^{+0.11}_{-0.15}$ & 
8.53$^{+0.23}_{-0.22}$ &  8.78 & -0.19$^{+0.17}_{-0.22}$ & -1.08$^{+0.29}_{-0.31}$ & -1.33 &     \\
XBSJ005032.3$-$521543 & 1.216 & --$^{ }_{ }$ & 3.86$^{+0.023}_{-0.024}$ & 
45.33$^{+0.21}_{-0.30}$ & 45.53$^{+0.20}_{-0.30}$ &
--$^{ }_{ }$ & 9.42$^{+0.11}_{-0.16}$ & 
9.42$^{+0.11}_{-0.16}$ &  9.55 & 0.36$^{+0.16}_{-0.20}$ & -1.42$^{+0.19}_{-0.26}$ & -1.55 &     \\
XBSJ010421.4$-$061418 & 0.520 & --$^{ }_{ }$ & 3.48$^{+0.061}_{-0.072}$ & 
43.40$^{+0.15}_{-0.22}$ & 43.67$^{+0.14}_{-0.23}$ &
--$^{ }_{ }$ & 7.49$^{+0.13}_{-0.18}$ & 
7.49$^{+0.13}_{-0.18}$ &  7.93 & -0.89$^{+0.07}_{-0.09}$ & -0.74$^{+0.15}_{-0.20}$ & -1.18 &   2 \\
XBSJ010432.8$-$583712 & 1.640 & --$^{ }_{ }$ & 4.03$^{+0.020}_{-0.021}$ & 
45.62$^{+0.14}_{-0.16}$ & 45.85$^{+0.14}_{-0.16}$ &
--$^{ }_{ }$ & 9.94$^{+0.08}_{-0.09}$ & 
9.94$^{+0.08}_{-0.09}$ & 10.06 & 0.82$^{+0.10}_{-0.09}$ & -1.48$^{+0.13}_{-0.13}$ & -1.60 &   2 \\
XBSJ010701.5$-$172748 & 0.890 & --$^{ }_{ }$ & 3.58$^{+0.038}_{-0.042}$ & 
45.16$^{+0.17}_{-0.16}$ & 45.42$^{+0.18}_{-0.15}$ &
--$^{ }_{ }$ & 8.78$^{+0.11}_{-0.11}$ & 
8.78$^{+0.11}_{-0.11}$ &  9.35 & 0.60$^{+0.15}_{-0.13}$ & -0.54$^{+0.19}_{-0.17}$ & -1.11 &     \\
XBSJ010747.2$-$172044 & 0.980 & --$^{ }_{ }$ & 3.67$^{+0.011}_{-0.011}$ & 
45.91$^{+0.11}_{-0.16}$ & 46.14$^{+0.11}_{-0.16}$ &
--$^{ }_{ }$ & 9.41$^{+0.06}_{-0.09}$ & 
9.41$^{+0.06}_{-0.09}$ &  9.81 & 0.97$^{+0.11}_{-0.14}$ & -0.80$^{+0.13}_{-0.17}$ & -1.20 &     \\
XBSJ012000.0$-$110429 & 0.351 & 3.21$^{+0.060}_{-0.073}$ & --$^{ }_{ }$ & 
43.67$^{+0.21}_{-0.34}$ & 43.93$^{+0.22}_{-0.34}$ &
7.16$^{+0.14}_{-0.20}$ & --$^{ }_{ }$ & 
7.16$^{+0.14}_{-0.20}$ &  7.87 & -0.84$^{+0.17}_{-0.22}$ & -0.36$^{+0.22}_{-0.30}$ & -1.07 &   2 \\
XBSJ012025.2$-$105441 & 1.338 & --$^{ }_{ }$ & 3.77$^{+0.015}_{-0.016}$ & 
46.05$^{+0.14}_{-0.16}$ & 46.28$^{+0.14}_{-0.16}$ &
--$^{ }_{ }$ & 9.68$^{+0.08}_{-0.08}$ & 
9.68$^{+0.08}_{-0.08}$ & 10.01 & 1.11$^{+0.14}_{-0.14}$ & -0.93$^{+0.16}_{-0.16}$ & -1.26 &     \\
XBSJ012119.9$-$110418 & 0.204 & 3.57$^{+0.020}_{-0.021}$ & --$^{ }_{ }$ & 
44.15$^{+0.14}_{-0.16}$ & 44.37$^{+0.15}_{-0.15}$ &
8.13$^{+0.08}_{-0.09}$ & --$^{ }_{ }$ & 
8.13$^{+0.08}_{-0.09}$ &  8.33 & -0.72$^{+0.12}_{-0.12}$ & -1.21$^{+0.14}_{-0.15}$ & -1.41 &     \\
XBSJ013204.9$-$400050 & 0.445 & 3.43$^{+0.057}_{-0.049}$ & 3.67$^{+0.013}_{-0.007}$ & 
44.56$^{+0.14}_{-0.16}$ & 44.75$^{+0.15}_{-0.15}$ &
8.05$^{+0.13}_{-0.12}$ & 8.55$^{+0.08}_{-0.07}$ & 
8.05$^{+0.13}_{-0.12}$ &  8.40 & -0.47$^{+0.13}_{-0.13}$ & -0.88$^{+0.18}_{-0.18}$ & -1.23 &     \\
XBSJ014251.5+133352 & 1.071 & --$^{ }_{ }$ & 3.90$^{+0.022}_{-0.023}$ & 
45.52$^{+0.12}_{-0.09}$ & 45.75$^{+0.12}_{-0.10}$ &
--$^{ }_{ }$ & 9.63$^{+0.07}_{-0.07}$ & 
9.63$^{+0.07}_{-0.07}$ &  9.77 & 0.61$^{+0.10}_{-0.09}$ & -1.38$^{+0.12}_{-0.11}$ & -1.53 &     \\
XBSJ015957.5+003309 & 0.310 & 3.21$^{+0.071}_{-0.074}$ & 3.79$^{+0.181}_{-0.181}$ & 
44.06$^{+0.11}_{-0.10}$ & 44.19$^{+0.11}_{-0.10}$ &
7.36$^{+0.13}_{-0.12}$ & 8.44$^{+0.36}_{-0.36}$ & 
7.36$^{+0.13}_{-0.12}$ &  7.95 & -0.80$^{+0.05}_{-0.03}$ & -0.52$^{+0.14}_{-0.12}$ & -1.11 &   2 \\
XBSJ020029.0+002846 & 0.174 & 3.53$^{+0.084}_{-0.106}$ & --$^{ }_{ }$ & 
43.35$^{+0.12}_{-0.09}$ & 43.48$^{+0.12}_{-0.09}$ &
7.65$^{+0.17}_{-0.20}$ & --$^{ }_{ }$ & 
7.65$^{+0.17}_{-0.20}$ &  7.74 & -1.61$^{+0.06}_{-0.05}$ & -1.62$^{+0.18}_{-0.21}$ & -1.71 & 1   \\
XBSJ021808.3$-$045845 & 0.712 & 3.56$^{+0.057}_{-0.067}$ & 3.82$^{+0.011}_{-0.011}$ & 
45.55$^{+0.11}_{-0.10}$ & 45.74$^{+0.12}_{-0.09}$ &
8.81$^{+0.11}_{-0.12}$ & 9.45$^{+0.06}_{-0.05}$ & 
9.45$^{+0.06}_{-0.05}$ &  9.63 & 0.53$^{+0.09}_{-0.08}$ & -1.28$^{+0.11}_{-0.09}$ & -1.46 &     \\
XBSJ021817.4$-$045113 & 1.080 & --$^{ }_{ }$ & 3.82$^{+0.014}_{-0.014}$ & 
45.19$^{+0.12}_{-0.18}$ & 45.38$^{+0.13}_{-0.17}$ &
--$^{ }_{ }$ & 9.23$^{+0.07}_{-0.09}$ & 
9.23$^{+0.07}_{-0.09}$ &  9.46 & 0.46$^{+0.05}_{-0.07}$ & -1.13$^{+0.09}_{-0.11}$ & -1.36 &     \\
XBSJ021820.6$-$050427 & 0.646 & 3.31$^{+0.079}_{-0.100}$ & 3.68$^{+0.026}_{-0.027}$ & 
44.87$^{+0.08}_{-0.15}$ & 45.07$^{+0.08}_{-0.15}$ &
7.96$^{+0.14}_{-0.18}$ & 8.76$^{+0.06}_{-0.10}$ & 
8.76$^{+0.06}_{-0.10}$ &  8.95 & -0.12$^{+0.06}_{-0.12}$ & -1.24$^{+0.08}_{-0.16}$ & -1.43 &   2 \\
XBSJ021923.2$-$045148 & 0.632 & 3.76$^{+0.016}_{-0.017}$ & 3.69$^{+0.018}_{-0.019}$ & 
44.92$^{+0.11}_{-0.10}$ & 45.11$^{+0.12}_{-0.09}$ &
8.88$^{+0.07}_{-0.06}$ & 8.81$^{+0.07}_{-0.05}$ & 
8.81$^{+0.07}_{-0.05}$ &  8.99 & -0.11$^{+0.10}_{-0.08}$ & -1.28$^{+0.12}_{-0.09}$ & -1.46 &     \\
XBSJ023459.7$-$294436 & 0.446 & 3.67$^{+0.025}_{-0.016}$ & 3.68$^{+0.003}_{-0.003}$ & 
45.19$^{+0.21}_{-0.22}$ & 45.42$^{+0.21}_{-0.22}$ &
8.84$^{+0.12}_{-0.11}$ & 8.98$^{+0.10}_{-0.11}$ & 
8.84$^{+0.12}_{-0.11}$ &  9.14 & 0.23$^{+0.20}_{-0.22}$ & -0.97$^{+0.23}_{-0.25}$ & -1.28 &     \\
XBSJ024200.9+000020 & 1.112 & --$^{ }_{ }$ & 3.95$^{+0.017}_{-0.018}$ & 
45.74$^{+0.09}_{-0.05}$ & 45.87$^{+0.09}_{-0.05}$ &
--$^{ }_{ }$ & 9.79$^{+0.06}_{-0.04}$ & 
9.79$^{+0.06}_{-0.04}$ &  9.89 & 0.57$^{+0.07}_{-0.04}$ & -1.58$^{+0.09}_{-0.06}$ & -1.68 &     \\
XBSJ024204.7+000814 & 0.383 & 3.69$^{+0.031}_{-0.032}$ & 3.53$^{+0.016}_{-0.016}$ & 
44.31$^{+0.11}_{-0.10}$ & 44.58$^{+0.11}_{-0.10}$ &
8.45$^{+0.08}_{-0.08}$ & 8.17$^{+0.06}_{-0.06}$ & 
8.45$^{+0.08}_{-0.08}$ &  8.69 & -0.31$^{+0.11}_{-0.09}$ & -1.12$^{+0.14}_{-0.12}$ & -1.36 &     \\
XBSJ024207.3+000037 & 0.385 & 3.74$^{+0.047}_{-0.041}$ & --$^{ }_{ }$ & 
44.07$^{+0.08}_{-0.10}$ & 44.29$^{+0.08}_{-0.09}$ &
8.42$^{+0.10}_{-0.10}$ & --$^{ }_{ }$ & 
8.42$^{+0.10}_{-0.10}$ &  8.52 & -0.79$^{+0.06}_{-0.07}$ & -1.57$^{+0.12}_{-0.12}$ & -1.67 &     \\
XBSJ025606.1+001635 & 0.629 & 3.26$^{+0.143}_{-0.153}$ & 3.62$^{+0.014}_{-0.015}$ & 
44.65$^{+0.11}_{-0.10}$ & 44.85$^{+0.11}_{-0.10}$ &
7.76$^{+0.25}_{-0.24}$ & 8.50$^{+0.06}_{-0.06}$ & 
7.76$^{+0.25}_{-0.24}$ &  8.39 & -0.34$^{+0.09}_{-0.07}$ & -0.46$^{+0.27}_{-0.25}$ & -1.09 &   2 \\
XBSJ031015.5$-$765131 & 1.187 & --$^{ }_{ }$ & 3.98$^{+0.026}_{-0.028}$ & 
45.94$^{+0.13}_{-0.18}$ & 46.14$^{+0.12}_{-0.18}$ &
--$^{ }_{ }$ & 10.02$^{+0.08}_{-0.10}$ & 
10.02$^{+0.08}_{-0.10}$ & 10.16 & 0.99$^{+0.09}_{-0.12}$ & -1.39$^{+0.12}_{-0.16}$ & -1.53 &     \\
XBSJ031311.7$-$765428 & 1.274 & --$^{ }_{ }$ & 3.78$^{+0.033}_{-0.035}$ & 
45.66$^{+0.20}_{-0.23}$ & 45.88$^{+0.21}_{-0.22}$ &
--$^{ }_{ }$ & 9.47$^{+0.13}_{-0.13}$ & 
9.47$^{+0.13}_{-0.13}$ &  9.74 & 0.78$^{+0.17}_{-0.16}$ & -1.05$^{+0.21}_{-0.21}$ & -1.32 &     \\
XBSJ033208.7$-$274735 & 0.544 & 4.17$^{+0.030}_{-0.029}$ & --$^{ }_{ }$ & 
44.72$^{+0.08}_{-0.18}$ & 44.85$^{+0.09}_{-0.18}$ &
9.60$^{+0.07}_{-0.11}$ & --$^{ }_{ }$ & 
9.60$^{+0.07}_{-0.11}$ &  9.62 & -0.45$^{+0.07}_{-0.13}$ & -2.41$^{+0.10}_{-0.17}$ & -2.43 &     \\
\hline
\end{tabular}
\end{table}
\end{landscape}
\newpage
\begin{landscape}
\addtocounter{table}{-1}
\begin{table}
\caption{continue}
\begin{tabular}{r r r r r r r r r r r r r r}
\hline\hline
name & z & LogFWHM & LogFWHM  & log$\lambda$L$_{\lambda}$ & Log$\lambda$L$_{\lambda}$ & LogM$_{BH}$ & LogM$_{BH}$ & logM$_{BH}$ & log$M_{BH}$ & Log$\dot{M}$ & LogL/L$_{Edd}$ & LogL/L$_{Edd}$& Flag\\
     &   & H$\beta$ & MgII$\lambda$2798\AA & 5100\AA & 3000\AA& H$\beta$&MgII$\lambda$2798\AA  & best & p$_{rad}$  & & & p$_{rad}$ & \\
(1) & (2) & (3) & (4) & (5) & (6) & (7) & (8) & (9) & (10) & (11) & (12) & (13) & (14) \\
\hline\hline
XBSJ033506.0$-$255619 & 1.430 & --$^{ }_{ }$ & 3.77$^{+0.019}_{-0.020}$ & 
46.32$^{+0.18}_{-0.22}$ & 46.52$^{+0.17}_{-0.23}$ &
--$^{ }_{ }$ & 9.83$^{+0.10}_{-0.12}$ & 
9.83$^{+0.10}_{-0.12}$ & 10.15 & 1.25$^{+0.17}_{-0.20}$ & -0.94$^{+0.20}_{-0.23}$ & -1.26 &     \\
XBSJ033851.4$-$352646 & 1.070 & --$^{ }_{ }$ & 3.87$^{+0.037}_{-0.040}$ & 
45.88$^{+0.21}_{-0.22}$ & 46.02$^{+0.20}_{-0.22}$ &
--$^{ }_{ }$ & 9.74$^{+0.12}_{-0.13}$ & 
9.74$^{+0.12}_{-0.13}$ &  9.87 & 0.66$^{+0.19}_{-0.20}$ & -1.44$^{+0.22}_{-0.24}$ & -1.57 &     \\
XBSJ033912.1$-$352813 & 0.466 & 3.70$^{+0.037}_{-0.036}$ & 3.89$^{+0.010}_{-0.010}$ & 
44.16$^{+0.17}_{-0.16}$ & 44.29$^{+0.17}_{-0.16}$ &
8.38$^{+0.11}_{-0.11}$ & 8.70$^{+0.08}_{-0.09}$ & 
8.70$^{+0.08}_{-0.09}$ &  8.75 & -0.85$^{+0.11}_{-0.07}$ & -1.91$^{+0.14}_{-0.11}$ & -1.96 &     \\
XBSJ041108.1$-$711341 & 0.923 & --$^{ }_{ }$ & 3.75$^{+0.066}_{-0.078}$ & 
45.07$^{+0.23}_{-0.31}$ & 45.29$^{+0.24}_{-0.30}$ &
--$^{ }_{ }$ & 9.04$^{+0.18}_{-0.21}$ & 
9.04$^{+0.18}_{-0.21}$ &  9.24 & 0.20$^{+0.19}_{-0.21}$ & -1.20$^{+0.26}_{-0.30}$ & -1.40 &   2 \\
XBSJ050446.3$-$283821 & 0.840 & 3.56$^{+0.171}_{-0.179}$ & 4.01$^{+0.007}_{-0.007}$ & 
44.34$^{+0.15}_{-0.16}$ & 44.54$^{+0.14}_{-0.16}$ &
8.20$^{+0.35}_{-0.36}$ & 9.09$^{+0.07}_{-0.08}$ & 
8.20$^{+0.35}_{-0.36}$ &  8.49 & -0.44$^{+0.08}_{-0.06}$ & -1.00$^{+0.36}_{-0.36}$ & -1.29 &     \\
XBSJ050501.8$-$284149 & 0.257 & 3.36$^{+0.029}_{-0.029}$ & --$^{ }_{ }$ & 
43.64$^{+0.18}_{-0.15}$ & 43.84$^{+0.17}_{-0.16}$ &
7.44$^{+0.11}_{-0.09}$ & --$^{ }_{ }$ & 
7.44$^{+0.11}_{-0.09}$ &  7.67 & -1.33$^{+0.14}_{-0.11}$ & -1.13$^{+0.18}_{-0.14}$ & -1.36 &     \\
XBSJ051651.9+794314 & 0.557 & 3.50$^{+0.070}_{-0.085}$ & 3.63$^{+0.030}_{-0.033}$ & 
45.42$^{+0.23}_{-0.22}$ & 45.65$^{+0.23}_{-0.22}$ &
8.63$^{+0.17}_{-0.19}$ & 9.01$^{+0.13}_{-0.12}$ & 
9.01$^{+0.13}_{-0.12}$ &  9.37 & 0.50$^{+0.21}_{-0.20}$ & -0.87$^{+0.25}_{-0.23}$ & -1.23 &     \\
XBSJ051955.5$-$455727 & 0.562 & 3.61$^{+0.041}_{-0.045}$ & 3.63$^{+0.021}_{-0.022}$ & 
44.95$^{+0.11}_{-0.16}$ & 44.84$^{+0.11}_{-0.15}$ &
8.60$^{+0.10}_{-0.11}$ & 8.51$^{+0.07}_{-0.08}$ & 
8.51$^{+0.07}_{-0.08}$ &  8.72 & -0.31$^{+0.08}_{-0.11}$ & -1.18$^{+0.11}_{-0.14}$ & -1.39 &     \\
XBSJ052022.0$-$252309 & 0.745 & 4.31$^{+0.083}_{-0.065}$ & --$^{ }_{ }$ & 
44.90$^{+0.22}_{-0.43}$ & 45.03$^{+0.22}_{-0.44}$ &
9.97$^{+0.20}_{-0.25}$ & --$^{ }_{ }$ & 
9.97$^{+0.20}_{-0.25}$ &  9.98 & -0.20$^{+0.16}_{-0.25}$ & -2.53$^{+0.26}_{-0.35}$ & -2.54 &     \\
XBSJ052144.1$-$251518 & 0.321 & 3.73$^{+0.070}_{-0.068}$ & --$^{ }_{ }$ & 
43.90$^{+0.11}_{-0.16}$ & 44.09$^{+0.12}_{-0.15}$ &
8.32$^{+0.15}_{-0.15}$ & --$^{ }_{ }$ & 
8.32$^{+0.15}_{-0.15}$ &  8.39 & -1.08$^{+0.09}_{-0.11}$ & -1.76$^{+0.17}_{-0.19}$ & -1.83 &   2 \\
XBSJ052543.6$-$334856 & 0.735 & 3.75$^{+0.067}_{-0.065}$ & 3.82$^{+0.008}_{-0.008}$ & 
44.86$^{+0.23}_{-0.22}$ & 45.06$^{+0.23}_{-0.22}$ &
8.84$^{+0.18}_{-0.17}$ & 9.03$^{+0.11}_{-0.11}$ & 
8.84$^{+0.18}_{-0.17}$ &  9.01 & -0.10$^{+0.18}_{-0.15}$ & -1.30$^{+0.25}_{-0.23}$ & -1.47 &     \\
XBSJ065214.1+743230 & 0.620 & 3.51$^{+0.051}_{-0.058}$ & 3.86$^{+0.054}_{-0.039}$ & 
45.25$^{+0.20}_{-0.22}$ & 45.48$^{+0.20}_{-0.22}$ &
8.55$^{+0.14}_{-0.15}$ & 9.39$^{+0.14}_{-0.14}$ & 
9.39$^{+0.14}_{-0.14}$ &  9.52 & 0.33$^{+0.18}_{-0.19}$ & -1.42$^{+0.23}_{-0.24}$ & -1.55 &     \\
XBSJ065400.0+742045 & 0.362 & 3.56$^{+0.036}_{-0.032}$ & 3.27$^{+0.042}_{-0.047}$ & 
44.41$^{+0.15}_{-0.15}$ & 44.61$^{+0.15}_{-0.15}$ &
8.24$^{+0.10}_{-0.10}$ & 7.67$^{+0.10}_{-0.11}$ & 
8.24$^{+0.10}_{-0.10}$ &  8.44 & -0.61$^{+0.12}_{-0.13}$ & -1.21$^{+0.16}_{-0.16}$ & -1.41 & 1   \\
XBSJ074202.7+742625 & 0.599 & 3.72$^{+0.062}_{-0.073}$ & 3.76$^{+0.011}_{-0.011}$ & 
44.16$^{+0.14}_{-0.16}$ & 44.40$^{+0.15}_{-0.15}$ &
8.42$^{+0.14}_{-0.16}$ & 8.51$^{+0.08}_{-0.08}$ & 
8.51$^{+0.08}_{-0.08}$ &  8.71 & -0.35$^{+0.07}_{-0.06}$ & -1.22$^{+0.11}_{-0.10}$ & -1.42 &   2 \\
XBSJ074352.0+744258 & 0.800 & --$^{ }_{ }$ & 3.72$^{+0.027}_{-0.028}$ & 
45.22$^{+0.11}_{-0.16}$ & 45.41$^{+0.12}_{-0.15}$ &
--$^{ }_{ }$ & 9.06$^{+0.08}_{-0.09}$ & 
9.06$^{+0.08}_{-0.09}$ &  9.26 & 0.21$^{+0.10}_{-0.12}$ & -1.21$^{+0.13}_{-0.15}$ & -1.41 &     \\
XBSJ080504.6+245156 & 0.980 & --$^{ }_{ }$ & 3.63$^{+0.073}_{-0.089}$ & 
44.43$^{+0.08}_{-0.10}$ & 44.63$^{+0.08}_{-0.10}$ &
--$^{ }_{ }$ & 8.39$^{+0.14}_{-0.17}$ & 
8.39$^{+0.14}_{-0.17}$ &  8.64 & -0.33$^{+0.03}_{-0.05}$ & -1.08$^{+0.14}_{-0.18}$ & -1.33 &   2 \\
XBSJ080608.1+244420 & 0.357 & 3.47$^{+0.028}_{-0.027}$ & 3.47$^{+0.017}_{-0.007}$ & 
44.62$^{+0.08}_{-0.10}$ & 44.84$^{+0.08}_{-0.09}$ &
8.15$^{+0.07}_{-0.07}$ & 8.21$^{+0.05}_{-0.05}$ & 
8.15$^{+0.07}_{-0.07}$ &  8.57 & -0.25$^{+0.06}_{-0.07}$ & -0.76$^{+0.09}_{-0.10}$ & -1.18 &     \\
XBSJ083049.8+524908 & 1.200 & --$^{ }_{ }$ & 3.63$^{+0.048}_{-0.055}$ & 
44.80$^{+0.08}_{-0.09}$ & 45.00$^{+0.08}_{-0.09}$ &
--$^{ }_{ }$ & 8.62$^{+0.10}_{-0.11}$ & 
8.62$^{+0.10}_{-0.11}$ &  9.01 & 0.16$^{+0.03}_{-0.03}$ & -0.82$^{+0.10}_{-0.11}$ & -1.21 &   2 \\
XBSJ083737.1+254751 & 0.080 & 4.20$^{+0.149}_{-0.031}$ & --$^{ }_{ }$ & 
43.53$^{+0.11}_{-0.10}$ & 43.76$^{+0.11}_{-0.10}$ &
9.07$^{+0.30}_{-0.08}$ & --$^{ }_{ }$ & 
9.07$^{+0.30}_{-0.08}$ &  9.08 & -1.34$^{+0.09}_{-0.07}$ & -2.77$^{+0.31}_{-0.11}$ & -2.78 &     \\
XBSJ083838.6+253616 & 0.601 & --$^{ }_{ }$ & 3.75$^{+0.042}_{-0.047}$ & 
44.92$^{+0.12}_{-0.09}$ & 45.15$^{+0.11}_{-0.10}$ &
--$^{ }_{ }$ & 8.96$^{+0.10}_{-0.11}$ & 
8.96$^{+0.10}_{-0.11}$ &  9.12 & -0.02$^{+0.11}_{-0.09}$ & -1.34$^{+0.15}_{-0.14}$ & -1.50 &   2 \\
XBSJ083905.9+255010 & 0.250 & 3.86$^{+0.096}_{-0.106}$ & --$^{ }_{ }$ & 
43.23$^{+0.14}_{-0.16}$ & 43.45$^{+0.15}_{-0.15}$ &
8.24$^{+0.20}_{-0.22}$ & --$^{ }_{ }$ & 
8.24$^{+0.20}_{-0.22}$ &  8.27 & -1.55$^{+0.10}_{-0.09}$ & -2.15$^{+0.22}_{-0.24}$ & -2.18 &     \\
XBSJ085530.7+585129 & 0.905 & --$^{ }_{ }$ & 3.66$^{+0.034}_{-0.037}$ & 
44.41$^{+0.08}_{-0.15}$ & 44.64$^{+0.08}_{-0.16}$ &
--$^{ }_{ }$ & 8.46$^{+0.07}_{-0.11}$ & 
8.46$^{+0.07}_{-0.11}$ &  8.77 & -0.14$^{+0.03}_{-0.05}$ & -0.96$^{+0.08}_{-0.12}$ & -1.27 &   2 \\
XBSJ094548.3$-$084824 & 1.748 & --$^{ }_{ }$ & 3.72$^{+0.010}_{-0.010}$ & 
46.31$^{+0.19}_{-0.34}$ & 46.55$^{+0.19}_{-0.34}$ &
--$^{ }_{ }$ & 9.77$^{+0.09}_{-0.17}$ & 
9.77$^{+0.09}_{-0.17}$ & 10.25 & 1.46$^{+0.19}_{-0.31}$ & -0.67$^{+0.21}_{-0.35}$ & -1.15 &     \\
XBSJ095054.5+393924 & 1.299 & --$^{ }_{ }$ & 3.58$^{+0.047}_{-0.053}$ & 
45.41$^{+0.11}_{-0.10}$ & 45.65$^{+0.12}_{-0.09}$ &
--$^{ }_{ }$ & 8.93$^{+0.11}_{-0.11}$ & 
8.93$^{+0.11}_{-0.11}$ &  9.40 & 0.61$^{+0.10}_{-0.08}$ & -0.68$^{+0.15}_{-0.14}$ & -1.16 &     \\
XBSJ095309.7+013558 & 0.477 & 3.87$^{+0.138}_{-0.206}$ & 3.61$^{+0.038}_{-0.042}$ & 
43.94$^{+0.11}_{-0.16}$ & 44.21$^{+0.11}_{-0.16}$ &
8.63$^{+0.28}_{-0.41}$ & 8.08$^{+0.10}_{-0.11}$ & 
8.08$^{+0.10}_{-0.11}$ &  8.37 & -0.57$^{+0.09}_{-0.12}$ & -1.01$^{+0.13}_{-0.16}$ & -1.30 &     \\
XBSJ095509.6+174124 & 1.290 & --$^{ }_{ }$ & 3.75$^{+0.053}_{-0.060}$ & 
45.13$^{+0.08}_{-0.10}$ & 45.40$^{+0.08}_{-0.10}$ &
--$^{ }_{ }$ & 9.12$^{+0.11}_{-0.13}$ & 
9.12$^{+0.11}_{-0.13}$ &  9.47 & 0.59$^{+0.07}_{-0.07}$ & -0.89$^{+0.13}_{-0.15}$ & -1.24 &   2 \\
XBSJ100100.0+252103 & 0.794 & --$^{ }_{ }$ & 3.68$^{+0.010}_{-0.011}$ & 
44.96$^{+0.11}_{-0.10}$ & 45.09$^{+0.12}_{-0.09}$ &
--$^{ }_{ }$ & 8.78$^{+0.06}_{-0.05}$ & 
8.78$^{+0.06}_{-0.05}$ &  8.95 & -0.15$^{+0.08}_{-0.07}$ & -1.29$^{+0.10}_{-0.09}$ & -1.46 &     \\
XBSJ100309.4+554135 & 0.673 & 3.88$^{+0.050}_{-0.051}$ & 3.71$^{+0.010}_{-0.010}$ & 
44.91$^{+0.08}_{-0.10}$ & 45.14$^{+0.08}_{-0.10}$ &
9.12$^{+0.11}_{-0.11}$ & 8.87$^{+0.05}_{-0.05}$ & 
8.87$^{+0.05}_{-0.05}$ &  9.06 & -0.01$^{+0.08}_{-0.08}$ & -1.23$^{+0.09}_{-0.09}$ & -1.42 &     \\
XBSJ100828.8+535408 & 0.384 & 3.56$^{+0.033}_{-0.033}$ & 3.85$^{+0.149}_{-0.123}$ & 
44.37$^{+0.08}_{-0.09}$ & 44.51$^{+0.08}_{-0.10}$ &
8.21$^{+0.07}_{-0.08}$ & 8.75$^{+0.30}_{-0.24}$ & 
8.75$^{+0.30}_{-0.24}$ &  8.80 & -0.82$^{+0.07}_{-0.08}$ & -1.93$^{+0.31}_{-0.25}$ & -1.98 &     \\
XBSJ100921.7+534926 & 0.387 & 3.63$^{+0.049}_{-0.046}$ & 3.74$^{+0.007}_{-0.007}$ & 
44.10$^{+0.12}_{-0.15}$ & 44.30$^{+0.11}_{-0.15}$ &
8.22$^{+0.12}_{-0.12}$ & 8.41$^{+0.05}_{-0.08}$ & 
8.22$^{+0.12}_{-0.12}$ &  8.36 & -0.83$^{+0.08}_{-0.10}$ & -1.41$^{+0.14}_{-0.16}$ & -1.55 &     \\
XBSJ100926.5+533426 & 1.718 & --$^{ }_{ }$ & 3.68$^{+0.075}_{-0.091}$ & 
45.80$^{+0.11}_{-0.10}$ & 45.99$^{+0.12}_{-0.10}$ &
--$^{ }_{ }$ & 9.34$^{+0.16}_{-0.19}$ & 
9.34$^{+0.16}_{-0.19}$ &  9.68 & 0.80$^{+0.09}_{-0.07}$ & -0.90$^{+0.18}_{-0.20}$ & -1.24 &   2 \\
XBSJ101506.0+520157 & 0.610 & 3.40$^{+0.050}_{-0.058}$ & 3.79$^{+0.028}_{-0.003}$ & 
44.73$^{+0.11}_{-0.10}$ & 44.97$^{+0.12}_{-0.09}$ &
8.07$^{+0.10}_{-0.11}$ & 8.92$^{+0.08}_{-0.04}$ & 
8.92$^{+0.08}_{-0.04}$ &  9.07 & -0.09$^{+0.11}_{-0.08}$ & -1.37$^{+0.14}_{-0.09}$ & -1.52 &   2 \\
XBSJ101838.0+411635 & 0.577 & 3.48$^{+0.132}_{-0.341}$ & 3.82$^{+0.014}_{-0.014}$ & 
44.43$^{+0.08}_{-0.10}$ & 44.67$^{+0.08}_{-0.09}$ &
8.09$^{+0.25}_{-0.54}$ & 8.79$^{+0.05}_{-0.06}$ & 
8.79$^{+0.05}_{-0.06}$ &  8.91 & -0.33$^{+0.07}_{-0.07}$ & -1.48$^{+0.09}_{-0.09}$ & -1.60 &     \\
XBSJ101850.5+411506 & 0.577 & 3.87$^{+0.186}_{-0.335}$ & 3.69$^{+0.005}_{-0.005}$ & 
45.06$^{+0.07}_{-0.10}$ & 45.25$^{+0.08}_{-0.09}$ &
9.17$^{+0.37}_{-0.65}$ & 8.89$^{+0.05}_{-0.04}$ & 
8.89$^{+0.05}_{-0.04}$ &  9.10 & 0.07$^{+0.07}_{-0.08}$ & -1.18$^{+0.09}_{-0.08}$ & -1.39 &     \\
XBSJ101922.6+412049 & 0.239 & 4.04$^{+0.033}_{-0.378}$ & --$^{ }_{ }$ & 
43.83$^{+0.11}_{-0.10}$ & 43.96$^{+0.11}_{-0.10}$ &
8.90$^{+0.08}_{-0.75}$ & --$^{ }_{ }$ & 
8.90$^{+0.08}_{-0.75}$ &  8.92 & -1.05$^{+0.05}_{-0.04}$ & -2.31$^{+0.09}_{-0.75}$ & -2.33 &     \\
\hline
\end{tabular}
\end{table}
\end{landscape}
\newpage
\begin{landscape}
\addtocounter{table}{-1}
\begin{table}
\caption{continue}
\begin{tabular}{r r r r r r r r r r r r r r}
\hline\hline
name & z & LogFWHM & LogFWHM  & log$\lambda$L$_{\lambda}$ & Log$\lambda$L$_{\lambda}$ & LogM$_{BH}$ & LogM$_{BH}$ & logM$_{BH}$ & log$M_{BH}$ & Log$\dot{M}$ & LogL/L$_{Edd}$ & LogL/L$_{Edd}$& Flag\\
     &   & H$\beta$ & MgII$\lambda$2798\AA & 5100\AA & 3000\AA& H$\beta$&MgII$\lambda$2798\AA  & best & p$_{rad}$  & & & p$_{rad}$ & \\
(1) & (2) & (3) & (4) & (5) & (6) & (7) & (8) & (9) & (10) & (11) & (12) & (13) & (14) \\
\hline\hline
XBSJ102412.3+042023 & 1.458 & --$^{ }_{ }$ & 3.68$^{+0.130}_{-0.189}$ & 
45.54$^{+0.11}_{-0.10}$ & 45.77$^{+0.11}_{-0.10}$ &
--$^{ }_{ }$ & 9.19$^{+0.26}_{-0.37}$ & 
9.19$^{+0.26}_{-0.37}$ &  9.53 & 0.64$^{+0.10}_{-0.07}$ & -0.91$^{+0.28}_{-0.38}$ & -1.25 &     \\
XBSJ103120.0+311404 & 1.190 & --$^{ }_{ }$ & 3.86$^{+0.020}_{-0.021}$ & 
45.08$^{+0.15}_{-0.10}$ & 45.31$^{+0.15}_{-0.10}$ &
--$^{ }_{ }$ & 9.27$^{+0.09}_{-0.06}$ & 
9.27$^{+0.09}_{-0.06}$ &  9.45 & 0.35$^{+0.09}_{-0.05}$ & -1.28$^{+0.13}_{-0.08}$ & -1.46 &   2 \\
XBSJ103154.1+310732 & 0.299 & 4.20$^{+0.132}_{-0.093}$ & --$^{ }_{ }$ & 
43.89$^{+0.08}_{-0.09}$ & 44.03$^{+0.07}_{-0.10}$ &
9.25$^{+0.26}_{-0.19}$ & --$^{ }_{ }$ & 
9.25$^{+0.26}_{-0.19}$ &  9.26 & -1.22$^{+0.06}_{-0.06}$ & -2.83$^{+0.27}_{-0.20}$ & -2.84 & 1   \\
XBSJ103909.4+205222 & 0.980 & --$^{ }_{ }$ & 3.71$^{+0.016}_{-0.017}$ & 
45.16$^{+0.11}_{-0.10}$ & 45.36$^{+0.11}_{-0.10}$ &
--$^{ }_{ }$ & 9.00$^{+0.07}_{-0.05}$ & 
9.00$^{+0.07}_{-0.05}$ &  9.24 & 0.24$^{+0.08}_{-0.06}$ & -1.12$^{+0.11}_{-0.08}$ & -1.36 &     \\
XBSJ103932.7+205426 & 0.237 & 3.64$^{+0.082}_{-0.066}$ & --$^{ }_{ }$ & 
43.66$^{+0.12}_{-0.09}$ & 43.80$^{+0.11}_{-0.10}$ &
8.02$^{+0.17}_{-0.13}$ & --$^{ }_{ }$ & 
8.02$^{+0.17}_{-0.13}$ &  8.09 & -1.36$^{+0.07}_{-0.05}$ & -1.74$^{+0.18}_{-0.14}$ & -1.81 &   2 \\
XBSJ103935.8+533036 & 0.229 & 3.82$^{+0.023}_{-0.024}$ & --$^{ }_{ }$ & 
44.32$^{+0.11}_{-0.15}$ & 44.21$^{+0.12}_{-0.15}$ &
8.70$^{+0.07}_{-0.09}$ & --$^{ }_{ }$ & 
8.70$^{+0.07}_{-0.09}$ &  8.74 & -0.99$^{+0.09}_{-0.12}$ & -2.05$^{+0.11}_{-0.15}$ & -2.09 &     \\
XBSJ104026.9+204542 & 0.465 & 3.73$^{+0.014}_{-0.032}$ & 3.63$^{+0.055}_{-0.063}$ & 
44.32$^{+0.08}_{-0.10}$ & 44.58$^{+0.08}_{-0.09}$ &
8.52$^{+0.05}_{-0.08}$ & 8.36$^{+0.11}_{-0.13}$ & 
8.52$^{+0.05}_{-0.08}$ &  8.87 & -0.01$^{+0.04}_{-0.04}$ & -0.89$^{+0.06}_{-0.09}$ & -1.24 &     \\
XBSJ104034.3+205110 & 0.670 & --$^{ }_{ }$ & 3.74$^{+0.029}_{-0.032}$ & 
45.04$^{+0.15}_{-0.16}$ & 45.28$^{+0.15}_{-0.15}$ &
--$^{ }_{ }$ & 9.02$^{+0.09}_{-0.10}$ & 
9.02$^{+0.09}_{-0.10}$ &  9.24 & 0.21$^{+0.13}_{-0.15}$ & -1.17$^{+0.16}_{-0.18}$ & -1.39 &   2 \\
XBSJ104509.3$-$012442 & 0.472 & 3.53$^{+0.020}_{-0.019}$ & 3.48$^{+0.011}_{-0.011}$ & 
44.08$^{+0.08}_{-0.09}$ & 44.28$^{+0.08}_{-0.10}$ &
8.00$^{+0.06}_{-0.05}$ & 7.86$^{+0.05}_{-0.05}$ & 
8.00$^{+0.06}_{-0.05}$ &  8.20 & -0.85$^{+0.05}_{-0.06}$ & -1.21$^{+0.08}_{-0.08}$ & -1.41 &     \\
XBSJ104912.8+330459 & 0.226 & 3.98$^{+0.105}_{-0.083}$ & --$^{ }_{ }$ & 
43.17$^{+0.11}_{-0.15}$ & 43.30$^{+0.11}_{-0.15}$ &
8.46$^{+0.21}_{-0.18}$ & --$^{ }_{ }$ & 
8.46$^{+0.21}_{-0.18}$ &  8.48 & -1.40$^{+0.02}_{-0.03}$ & -2.22$^{+0.21}_{-0.18}$ & -2.24 & 1   \\
XBSJ105014.9+331013 & 1.012 & --$^{ }_{ }$ & 3.88$^{+0.058}_{-0.024}$ & 
45.78$^{+0.11}_{-0.14}$ & 45.98$^{+0.11}_{-0.14}$ &
--$^{ }_{ }$ & 9.72$^{+0.13}_{-0.09}$ & 
9.72$^{+0.13}_{-0.09}$ &  9.87 & 0.71$^{+0.10}_{-0.13}$ & -1.37$^{+0.16}_{-0.16}$ & -1.52 &     \\
XBSJ105239.7+572431 & 1.113 & --$^{ }_{ }$ & 3.73$^{+0.014}_{-0.015}$ & 
45.87$^{+0.08}_{-0.10}$ & 46.07$^{+0.08}_{-0.10}$ &
--$^{ }_{ }$ & 9.48$^{+0.05}_{-0.06}$ & 
9.48$^{+0.05}_{-0.06}$ &  9.76 & 0.82$^{+0.07}_{-0.09}$ & -1.02$^{+0.09}_{-0.11}$ & -1.30 &   2 \\
XBSJ105316.9+573551 & 1.204 & --$^{ }_{ }$ & 3.55$^{+0.056}_{-0.064}$ & 
45.38$^{+0.08}_{-0.09}$ & 45.58$^{+0.08}_{-0.10}$ &
--$^{ }_{ }$ & 8.82$^{+0.12}_{-0.14}$ & 
8.82$^{+0.12}_{-0.14}$ &  9.31 & 0.53$^{+0.05}_{-0.05}$ & -0.65$^{+0.13}_{-0.15}$ & -1.15 &     \\
XBSJ105624.2$-$033522 & 0.635 & --$^{ }_{ }$ & 3.69$^{+0.012}_{-0.012}$ & 
44.82$^{+0.07}_{-0.10}$ & 45.01$^{+0.08}_{-0.09}$ &
--$^{ }_{ }$ & 8.75$^{+0.05}_{-0.05}$ & 
8.75$^{+0.05}_{-0.05}$ &  8.92 & -0.20$^{+0.07}_{-0.08}$ & -1.31$^{+0.09}_{-0.09}$ & -1.48 &     \\
XBSJ110652.0$-$182738 & 1.435 & --$^{ }_{ }$ & 3.73$^{+0.096}_{-0.125}$ & 
45.29$^{+0.34}_{-0.40}$ & 45.52$^{+0.34}_{-0.40}$ &
--$^{ }_{ }$ & 9.15$^{+0.25}_{-0.31}$ & 
9.15$^{+0.25}_{-0.31}$ &  9.47 & 0.57$^{+0.22}_{-0.18}$ & -0.94$^{+0.33}_{-0.36}$ & -1.26 &   2 \\
XBSJ112022.3+125252 & 0.406 & 3.46$^{+0.041}_{-0.042}$ & 3.61$^{+0.021}_{-0.023}$ & 
44.26$^{+0.08}_{-0.09}$ & 44.49$^{+0.08}_{-0.10}$ &
7.96$^{+0.08}_{-0.08}$ & 8.26$^{+0.06}_{-0.06}$ & 
8.26$^{+0.06}_{-0.06}$ &  8.47 & -0.57$^{+0.06}_{-0.06}$ & -1.19$^{+0.08}_{-0.08}$ & -1.40 &     \\
XBSJ112046.7+125429 & 0.382 & 3.82$^{+0.037}_{-0.041}$ & 3.71$^{+0.035}_{-0.038}$ & 
44.25$^{+0.08}_{-0.09}$ & 44.48$^{+0.08}_{-0.10}$ &
8.67$^{+0.08}_{-0.09}$ & 8.45$^{+0.08}_{-0.09}$ & 
8.67$^{+0.08}_{-0.09}$ &  8.76 & -0.59$^{+0.06}_{-0.07}$ & -1.62$^{+0.10}_{-0.11}$ & -1.71 & 1 2 \\
XBSJ113106.9+312518 & 1.482 & --$^{ }_{ }$ & 3.59$^{+0.019}_{-0.020}$ & 
45.71$^{+0.14}_{-0.16}$ & 45.94$^{+0.14}_{-0.16}$ &
--$^{ }_{ }$ & 9.13$^{+0.08}_{-0.09}$ & 
9.13$^{+0.08}_{-0.09}$ &  9.63 & 0.85$^{+0.11}_{-0.12}$ & -0.64$^{+0.14}_{-0.15}$ & -1.14 &     \\
XBSJ115317.9+364712 & 0.725 & --$^{ }_{ }$ & 3.53$^{+0.024}_{-0.026}$ & 
44.53$^{+0.15}_{-0.15}$ & 44.80$^{+0.14}_{-0.16}$ &
--$^{ }_{ }$ & 8.30$^{+0.08}_{-0.09}$ & 
8.30$^{+0.08}_{-0.09}$ &  8.75 & -0.05$^{+0.13}_{-0.14}$ & -0.71$^{+0.15}_{-0.17}$ & -1.17 &     \\
XBSJ120359.1+443715 & 0.641 & --$^{ }_{ }$ & 3.72$^{+0.014}_{-0.015}$ & 
44.76$^{+0.11}_{-0.10}$ & 44.95$^{+0.12}_{-0.09}$ &
--$^{ }_{ }$ & 8.77$^{+0.06}_{-0.06}$ & 
8.77$^{+0.06}_{-0.06}$ &  8.89 & -0.34$^{+0.11}_{-0.10}$ & -1.47$^{+0.13}_{-0.12}$ & -1.59 &     \\
XBSJ120413.7+443149 & 0.492 & --$^{ }_{ }$ & 3.67$^{+0.013}_{-0.014}$ & 
43.93$^{+0.08}_{-0.09}$ & 44.20$^{+0.08}_{-0.10}$ &
--$^{ }_{ }$ & 8.21$^{+0.04}_{-0.06}$ & 
8.21$^{+0.04}_{-0.06}$ &  8.39 & -0.70$^{+0.08}_{-0.09}$ & -1.27$^{+0.09}_{-0.11}$ & -1.45 &     \\
XBSJ123036.2+642531 & 0.744 & --$^{ }_{ }$ & 3.56$^{+0.036}_{-0.039}$ & 
44.67$^{+0.12}_{-0.09}$ & 44.80$^{+0.12}_{-0.09}$ &
--$^{ }_{ }$ & 8.36$^{+0.09}_{-0.08}$ & 
8.36$^{+0.09}_{-0.08}$ &  8.58 & -0.44$^{+0.08}_{-0.07}$ & -1.16$^{+0.12}_{-0.11}$ & -1.38 &   2 \\
XBSJ123116.5+641115 & 0.454 & 4.18$^{+0.086}_{-0.062}$ & --$^{ }_{ }$ & 
43.89$^{+0.08}_{-0.10}$ & 44.03$^{+0.07}_{-0.10}$ &
9.21$^{+0.18}_{-0.13}$ & --$^{ }_{ }$ & 
9.21$^{+0.18}_{-0.13}$ &  9.22 & -1.07$^{+0.05}_{-0.04}$ & -2.64$^{+0.19}_{-0.14}$ & -2.65 & 1 2 \\
XBSJ123218.5+640311 & 1.013 & --$^{ }_{ }$ & 3.51$^{+0.130}_{-0.198}$ & 
44.79$^{+0.14}_{-0.16}$ & 44.98$^{+0.15}_{-0.15}$ &
--$^{ }_{ }$ & 8.36$^{+0.25}_{-0.34}$ & 
8.36$^{+0.25}_{-0.34}$ &  8.78 & -0.04$^{+0.08}_{-0.07}$ & -0.76$^{+0.26}_{-0.35}$ & -1.18 &   2 \\
XBSJ123759.6+621102 & 0.910 & --$^{ }_{ }$ & 3.71$^{+0.007}_{-0.007}$ & 
45.41$^{+0.08}_{-0.09}$ & 45.61$^{+0.08}_{-0.10}$ &
--$^{ }_{ }$ & 9.16$^{+0.05}_{-0.05}$ & 
9.16$^{+0.05}_{-0.05}$ &  9.40 & 0.40$^{+0.06}_{-0.08}$ & -1.12$^{+0.08}_{-0.09}$ & -1.36 &     \\
XBSJ123800.9+621338 & 0.440 & 3.62$^{+0.044}_{-0.044}$ & 3.79$^{+0.002}_{-0.002}$ & 
44.57$^{+0.08}_{-0.09}$ & 44.77$^{+0.08}_{-0.09}$ &
8.44$^{+0.09}_{-0.10}$ & 8.80$^{+0.04}_{-0.04}$ & 
8.44$^{+0.09}_{-0.10}$ &  8.62 & -0.48$^{+0.07}_{-0.08}$ & -1.28$^{+0.11}_{-0.13}$ & -1.46 &     \\
XBSJ124214.1$-$112512 & 0.820 & --$^{ }_{ }$ & 3.65$^{+0.014}_{-0.014}$ & 
45.25$^{+0.13}_{-0.11}$ & 45.38$^{+0.13}_{-0.11}$ &
--$^{ }_{ }$ & 8.89$^{+0.07}_{-0.06}$ & 
8.89$^{+0.07}_{-0.06}$ &  9.12 & 0.12$^{+0.09}_{-0.08}$ & -1.13$^{+0.11}_{-0.10}$ & -1.36 &     \\
XBSJ124557.6+022659 & 0.708 & 3.58$^{+0.075}_{-0.075}$ & 3.69$^{+0.028}_{-0.029}$ & 
44.70$^{+0.15}_{-0.15}$ & 44.97$^{+0.15}_{-0.16}$ &
8.41$^{+0.17}_{-0.16}$ & 8.71$^{+0.09}_{-0.10}$ & 
8.71$^{+0.09}_{-0.10}$ &  9.04 & 0.15$^{+0.12}_{-0.13}$ & -0.92$^{+0.15}_{-0.16}$ & -1.25 &     \\
XBSJ124607.6+022153 & 0.491 & 3.64$^{+0.048}_{-0.047}$ & 3.58$^{+0.011}_{-0.012}$ & 
44.41$^{+0.08}_{-0.09}$ & 44.64$^{+0.08}_{-0.10}$ &
8.40$^{+0.10}_{-0.10}$ & 8.30$^{+0.04}_{-0.06}$ & 
8.40$^{+0.10}_{-0.10}$ &  8.61 & -0.42$^{+0.06}_{-0.07}$ & -1.18$^{+0.12}_{-0.12}$ & -1.39 & 1   \\
XBSJ124641.8+022412 & 0.934 & --$^{ }_{ }$ & 3.55$^{+0.005}_{-0.005}$ & 
45.89$^{+0.04}_{-0.10}$ & 46.03$^{+0.04}_{-0.10}$ &
--$^{ }_{ }$ & 9.11$^{+0.02}_{-0.06}$ & 
9.11$^{+0.02}_{-0.06}$ &  9.53 & 0.70$^{+0.03}_{-0.08}$ & -0.77$^{+0.04}_{-0.10}$ & -1.19 &     \\
XBSJ124647.9+020955 & 1.074 & --$^{ }_{ }$ & 3.81$^{+0.046}_{-0.052}$ & 
45.12$^{+0.15}_{-0.15}$ & 45.35$^{+0.15}_{-0.15}$ &
--$^{ }_{ }$ & 9.20$^{+0.12}_{-0.13}$ & 
9.20$^{+0.12}_{-0.13}$ &  9.36 & 0.22$^{+0.13}_{-0.12}$ & -1.34$^{+0.18}_{-0.18}$ & -1.50 &     \\
XBSJ124914.6$-$060910 & 1.627 & --$^{ }_{ }$ & 3.75$^{+0.025}_{-0.027}$ & 
45.74$^{+0.08}_{-0.09}$ & 45.97$^{+0.08}_{-0.10}$ &
--$^{ }_{ }$ & 9.46$^{+0.06}_{-0.07}$ & 
9.46$^{+0.06}_{-0.07}$ &  9.75 & 0.83$^{+0.07}_{-0.09}$ & -0.99$^{+0.09}_{-0.11}$ & -1.29 &   2 \\
XBSJ124949.4$-$060722 & 1.053 & --$^{ }_{ }$ & 3.42$^{+0.017}_{-0.017}$ & 
45.35$^{+0.08}_{-0.10}$ & 45.55$^{+0.08}_{-0.10}$ &
--$^{ }_{ }$ & 8.53$^{+0.05}_{-0.06}$ & 
8.53$^{+0.05}_{-0.06}$ &  9.10 & 0.34$^{+0.06}_{-0.08}$ & -0.55$^{+0.08}_{-0.10}$ & -1.12 &     \\
XBSJ130619.7$-$233857 & 0.351 & 3.86$^{+0.035}_{-0.033}$ & --$^{ }_{ }$ & 
44.49$^{+0.15}_{-0.15}$ & 44.72$^{+0.15}_{-0.15}$ &
8.87$^{+0.11}_{-0.10}$ & --$^{ }_{ }$ & 
8.87$^{+0.11}_{-0.10}$ &  8.96 & -0.39$^{+0.12}_{-0.12}$ & -1.62$^{+0.16}_{-0.16}$ & -1.71 &     \\
XBSJ130658.1$-$234849 & 0.375 & 3.63$^{+0.039}_{-0.038}$ & --$^{ }_{ }$ & 
44.44$^{+0.17}_{-0.23}$ & 44.57$^{+0.17}_{-0.22}$ &
8.39$^{+0.12}_{-0.13}$ & --$^{ }_{ }$ & 
8.39$^{+0.12}_{-0.13}$ &  8.51 & -0.73$^{+0.15}_{-0.17}$ & -1.48$^{+0.19}_{-0.21}$ & -1.60 &     \\
\hline
\end{tabular}
\end{table}
\end{landscape}
\newpage
\begin{landscape}
\addtocounter{table}{-1}
\begin{table}
\caption{continue}
\begin{tabular}{r r r r r r r r r r r r r r}
\hline\hline
name & z & LogFWHM & LogFWHM  & log$\lambda$L$_{\lambda}$ & Log$\lambda$L$_{\lambda}$ & LogM$_{BH}$ & LogM$_{BH}$ & logM$_{BH}$ & log$M_{BH}$ & Log$\dot{M}$ & LogL/L$_{Edd}$ & LogL/L$_{Edd}$& Flag\\
     &   & H$\beta$ & MgII$\lambda$2798\AA & 5100\AA & 3000\AA& H$\beta$&MgII$\lambda$2798\AA  & best & p$_{rad}$  & & & p$_{rad}$ & \\
(1) & (2) & (3) & (4) & (5) & (6) & (7) & (8) & (9) & (10) & (11) & (12) & (13) & (14) \\
\hline\hline
XBSJ132038.0+341124 & 0.065 & 3.56$^{+0.042}_{-0.041}$ & --$^{ }_{ }$ & 
43.41$^{+0.12}_{-0.09}$ & 43.55$^{+0.11}_{-0.10}$ &
7.73$^{+0.10}_{-0.08}$ & --$^{ }_{ }$ & 
7.73$^{+0.10}_{-0.08}$ &  7.78 & -1.77$^{+0.09}_{-0.08}$ & -1.86$^{+0.13}_{-0.11}$ & -1.91 &     \\
XBSJ132101.6+340656 & 0.335 & 3.75$^{+0.028}_{-0.030}$ & 3.87$^{+0.059}_{-0.113}$ & 
44.17$^{+0.08}_{-0.10}$ & 44.43$^{+0.08}_{-0.09}$ &
8.49$^{+0.07}_{-0.08}$ & 8.75$^{+0.12}_{-0.23}$ & 
8.49$^{+0.07}_{-0.08}$ &  8.68 & -0.39$^{+0.06}_{-0.09}$ & -1.24$^{+0.09}_{-0.12}$ & -1.43 &     \\
XBSJ133807.5+242411 & 0.631 & 3.36$^{+0.023}_{-0.024}$ & 3.65$^{+0.010}_{-0.010}$ & 
45.24$^{+0.08}_{-0.09}$ & 45.44$^{+0.08}_{-0.10}$ &
8.24$^{+0.06}_{-0.06}$ & 8.93$^{+0.04}_{-0.06}$ & 
8.93$^{+0.04}_{-0.06}$ &  9.17 & 0.18$^{+0.07}_{-0.09}$ & -1.11$^{+0.08}_{-0.11}$ & -1.35 &     \\
XBSJ134749.9+582111 & 0.646 & 3.94$^{+0.030}_{-0.028}$ & 3.81$^{+0.004}_{-0.004}$ & 
45.74$^{+0.08}_{-0.10}$ & 45.96$^{+0.08}_{-0.09}$ &
9.65$^{+0.07}_{-0.07}$ & 9.58$^{+0.04}_{-0.05}$ & 
9.65$^{+0.07}_{-0.07}$ &  9.87 & 0.84$^{+0.06}_{-0.08}$ & -1.17$^{+0.09}_{-0.11}$ & -1.39 &     \\
XBSJ140102.0$-$111224 & 0.037 & 3.58$^{+0.489}_{--0.418}$ & --$^{ }_{ }$ & 
43.26$^{+0.09}_{-0.11}$ & 43.16$^{+0.09}_{-0.11}$ &
7.71$^{+0.96}_{-0.82}$ & --$^{ }_{ }$ & 
7.71$^{+0.96}_{-0.82}$ &  7.74 & -2.06$^{+0.07}_{-0.09}$ & -2.13$^{+0.96}_{-0.82}$ & -2.16 &     \\
XBSJ140113.4+024016 & 0.631 & 3.54$^{+0.220}_{-0.111}$ & 3.66$^{+0.038}_{-0.011}$ & 
43.58$^{+0.12}_{-0.15}$ & 43.85$^{+0.11}_{-0.16}$ &
7.77$^{+0.44}_{-0.22}$ & 7.97$^{+0.09}_{-0.08}$ & 
7.77$^{+0.44}_{-0.22}$ &  8.08 & -0.82$^{+0.07}_{-0.09}$ & -0.95$^{+0.45}_{-0.24}$ & -1.27 &   2 \\
XBSJ140127.7+025605 & 0.265 & 3.51$^{+0.017}_{-0.016}$ & --$^{ }_{ }$ & 
44.30$^{+0.08}_{-0.16}$ & 44.43$^{+0.08}_{-0.16}$ &
8.08$^{+0.05}_{-0.09}$ & --$^{ }_{ }$ & 
8.08$^{+0.05}_{-0.09}$ &  8.36 & -0.59$^{+0.03}_{-0.06}$ & -1.03$^{+0.06}_{-0.11}$ & -1.31 &     \\
XBSJ140921.1+261336 & 1.100 & --$^{ }_{ }$ & 3.62$^{+0.035}_{-0.039}$ & 
45.56$^{+0.17}_{-0.16}$ & 45.82$^{+0.18}_{-0.15}$ &
--$^{ }_{ }$ & 9.10$^{+0.12}_{-0.10}$ & 
9.10$^{+0.12}_{-0.10}$ &  9.74 & 1.01$^{+0.15}_{-0.12}$ & -0.45$^{+0.19}_{-0.16}$ & -1.09 &   2 \\
XBSJ141531.5+113156 & 0.257 & 4.09$^{+0.081}_{-0.074}$ & --$^{ }_{ }$ & 
44.08$^{+0.08}_{-0.10}$ & 43.97$^{+0.08}_{-0.09}$ &
9.13$^{+0.17}_{-0.15}$ & --$^{ }_{ }$ & 
9.13$^{+0.17}_{-0.15}$ &  9.14 & -1.06$^{+0.05}_{-0.05}$ & -2.55$^{+0.18}_{-0.16}$ & -2.56 &     \\
XBSJ144937.5+090826 & 1.260 & --$^{ }_{ }$ & 3.86$^{+0.017}_{-0.017}$ & 
45.47$^{+0.11}_{-0.10}$ & 45.66$^{+0.12}_{-0.09}$ &
--$^{ }_{ }$ & 9.50$^{+0.07}_{-0.06}$ & 
9.50$^{+0.07}_{-0.06}$ &  9.67 & 0.56$^{+0.08}_{-0.06}$ & -1.30$^{+0.11}_{-0.08}$ & -1.47 &     \\
XBSJ150428.3+101856 & 1.010 & --$^{ }_{ }$ & 3.73$^{+0.006}_{-0.007}$ & 
45.90$^{+0.11}_{-0.10}$ & 46.13$^{+0.11}_{-0.10}$ &
--$^{ }_{ }$ & 9.52$^{+0.05}_{-0.06}$ & 
9.52$^{+0.05}_{-0.06}$ &  9.85 & 0.95$^{+0.11}_{-0.09}$ & -0.93$^{+0.12}_{-0.11}$ & -1.26 &     \\
XBSJ151815.0+060851 & 1.294 & --$^{ }_{ }$ & 3.96$^{+0.011}_{-0.011}$ & 
45.42$^{+0.13}_{-0.20}$ & 45.65$^{+0.13}_{-0.20}$ &
--$^{ }_{ }$ & 9.68$^{+0.07}_{-0.10}$ & 
9.68$^{+0.07}_{-0.10}$ &  9.80 & 0.55$^{+0.11}_{-0.15}$ & -1.49$^{+0.13}_{-0.18}$ & -1.61 &     \\
XBSJ153205.7$-$082952 & 1.239 & --$^{ }_{ }$ & 3.78$^{+0.049}_{-0.055}$ & 
45.42$^{+0.15}_{-0.15}$ & 45.69$^{+0.14}_{-0.16}$ &
--$^{ }_{ }$ & 9.36$^{+0.11}_{-0.14}$ & 
9.36$^{+0.11}_{-0.14}$ &  9.71 & 0.83$^{+0.13}_{-0.14}$ & -0.89$^{+0.17}_{-0.20}$ & -1.24 &     \\
XBSJ153419.0+011808 & 1.283 & --$^{ }_{ }$ & 3.87$^{+0.020}_{-0.021}$ & 
45.55$^{+0.15}_{-0.15}$ & 45.82$^{+0.14}_{-0.16}$ &
--$^{ }_{ }$ & 9.61$^{+0.08}_{-0.09}$ & 
9.61$^{+0.08}_{-0.09}$ &  9.90 & 0.97$^{+0.13}_{-0.14}$ & -1.00$^{+0.15}_{-0.17}$ & -1.29 &     \\
XBSJ153456.1+013033 & 0.310 & 3.51$^{+0.037}_{-0.041}$ & 3.79$^{+0.013}_{-0.014}$ & 
44.70$^{+0.21}_{-0.15}$ & 44.95$^{+0.20}_{-0.16}$ &
8.27$^{+0.13}_{-0.11}$ & 8.90$^{+0.10}_{-0.08}$ & 
8.90$^{+0.10}_{-0.08}$ &  9.04 & -0.13$^{+0.19}_{-0.15}$ & -1.39$^{+0.21}_{-0.17}$ & -1.53 &     \\
XBSJ160706.6+075709 & 0.233 & 3.42$^{+0.055}_{-0.059}$ & --$^{ }_{ }$ & 
43.88$^{+0.08}_{-0.09}$ & 44.02$^{+0.08}_{-0.10}$ &
7.70$^{+0.10}_{-0.11}$ & --$^{ }_{ }$ & 
7.70$^{+0.10}_{-0.11}$ &  7.87 & -1.24$^{+0.06}_{-0.07}$ & -1.30$^{+0.12}_{-0.13}$ & -1.47 &     \\
XBSJ160731.5+081202 & 0.226 & 3.18$^{+0.059}_{-0.079}$ & --$^{ }_{ }$ & 
43.44$^{+0.12}_{-0.09}$ & 43.71$^{+0.11}_{-0.10}$ &
6.99$^{+0.09}_{-0.11}$ & --$^{ }_{ }$ & 
6.99$^{+0.09}_{-0.11}$ &  7.64 & -1.09$^{+0.09}_{-0.08}$ & -0.44$^{+0.13}_{-0.14}$ & -1.09 &     \\
XBSJ161615.1+121353 & 0.843 & --$^{ }_{ }$ & 3.90$^{+0.091}_{-0.104}$ & 
44.35$^{+0.08}_{-0.16}$ & 44.48$^{+0.08}_{-0.16}$ &
--$^{ }_{ }$ & 8.84$^{+0.19}_{-0.22}$ & 
8.84$^{+0.19}_{-0.22}$ &  8.92 & -0.49$^{+0.03}_{-0.05}$ & -1.69$^{+0.19}_{-0.23}$ & -1.77 &   2 \\
XBSJ161825.4+124145 & 0.396 & 3.53$^{+0.125}_{-0.127}$ & --$^{ }_{ }$ & 
44.22$^{+0.15}_{-0.09}$ & 44.42$^{+0.15}_{-0.10}$ &
8.08$^{+0.25}_{-0.24}$ & --$^{ }_{ }$ & 
8.08$^{+0.25}_{-0.24}$ &  8.28 & -0.78$^{+0.12}_{-0.08}$ & -1.22$^{+0.28}_{-0.25}$ & -1.42 &     \\
XBSJ165406.6+142123 & 0.641 & --$^{ }_{ }$ & 3.68$^{+0.030}_{-0.032}$ & 
45.40$^{+0.15}_{-0.16}$ & 45.29$^{+0.15}_{-0.15}$ &
--$^{ }_{ }$ & 8.90$^{+0.09}_{-0.10}$ & 
8.90$^{+0.09}_{-0.10}$ &  9.10 & 0.04$^{+0.13}_{-0.13}$ & -1.22$^{+0.16}_{-0.16}$ & -1.42 &     \\
XBSJ165425.3+142159 & 0.178 & 3.39$^{+0.141}_{-0.228}$ & --$^{ }_{ }$ & 
43.83$^{+0.12}_{-0.09}$ & 43.97$^{+0.11}_{-0.10}$ &
7.61$^{+0.26}_{-0.36}$ & --$^{ }_{ }$ & 
7.61$^{+0.26}_{-0.36}$ &  7.90 & -1.02$^{+0.04}_{-0.04}$ & -0.99$^{+0.26}_{-0.36}$ & -1.29 &   2 \\
XBSJ165448.5+141311 & 0.320 & --$^{ }_{ }$ & 4.01$^{+0.014}_{-0.014}$ & 
43.77$^{+0.08}_{-0.10}$ & 43.97$^{+0.08}_{-0.10}$ &
--$^{ }_{ }$ & 8.75$^{+0.05}_{-0.06}$ & 
8.75$^{+0.05}_{-0.06}$ &  8.81 & -0.68$^{+0.02}_{-0.02}$ & -1.79$^{+0.05}_{-0.06}$ & -1.85 &   2 \\
XBSJ185518.7$-$462504 & 0.788 & --$^{ }_{ }$ & 3.66$^{+0.021}_{-0.023}$ & 
45.83$^{+0.17}_{-0.23}$ & 45.96$^{+0.18}_{-0.22}$ &
--$^{ }_{ }$ & 9.28$^{+0.10}_{-0.12}$ & 
9.28$^{+0.10}_{-0.12}$ &  9.57 & 0.63$^{+0.16}_{-0.18}$ & -1.01$^{+0.19}_{-0.22}$ & -1.30 &     \\
XBSJ185613.7$-$462239 & 0.768 & --$^{ }_{ }$ & 3.63$^{+0.074}_{-0.091}$ & 
44.87$^{+0.15}_{-0.16}$ & 45.07$^{+0.14}_{-0.16}$ &
--$^{ }_{ }$ & 8.67$^{+0.16}_{-0.19}$ & 
8.67$^{+0.16}_{-0.19}$ &  8.96 & 0.02$^{+0.09}_{-0.08}$ & -1.01$^{+0.18}_{-0.21}$ & -1.30 &     \\
XBSJ204159.2$-$321439 & 0.738 & 3.47$^{+0.105}_{-0.142}$ & 3.60$^{+0.018}_{-0.019}$ & 
44.79$^{+0.15}_{-0.15}$ & 44.99$^{+0.15}_{-0.16}$ &
8.25$^{+0.21}_{-0.26}$ & 8.55$^{+0.09}_{-0.08}$ & 
8.55$^{+0.09}_{-0.08}$ &  8.83 & -0.11$^{+0.10}_{-0.10}$ & -1.02$^{+0.13}_{-0.13}$ & -1.30 &     \\
XBSJ204204.1$-$321601 & 0.384 & 3.87$^{+0.073}_{-0.088}$ & 3.85$^{+0.047}_{-0.052}$ & 
43.82$^{+0.17}_{-0.16}$ & 44.08$^{+0.18}_{-0.16}$ &
8.56$^{+0.16}_{-0.19}$ & 8.48$^{+0.13}_{-0.13}$ & 
8.48$^{+0.13}_{-0.13}$ &  8.57 & -0.77$^{+0.16}_{-0.13}$ & -1.61$^{+0.21}_{-0.18}$ & -1.70 &     \\
XBSJ204208.2$-$323523 & 1.184 & --$^{ }_{ }$ & 3.75$^{+0.021}_{-0.028}$ & 
44.88$^{+0.17}_{-0.30}$ & 45.12$^{+0.18}_{-0.30}$ &
--$^{ }_{ }$ & 8.93$^{+0.09}_{-0.16}$ & 
8.93$^{+0.09}_{-0.16}$ &  9.15 & 0.13$^{+0.14}_{-0.21}$ & -1.16$^{+0.17}_{-0.26}$ & -1.38 &   2 \\
XBSJ205635.7$-$044717 & 0.217 & 3.39$^{+0.030}_{-0.030}$ & --$^{ }_{ }$ & 
43.84$^{+0.15}_{-0.15}$ & 44.07$^{+0.15}_{-0.15}$ &
7.60$^{+0.10}_{-0.09}$ & --$^{ }_{ }$ & 
7.60$^{+0.10}_{-0.09}$ &  7.90 & -1.01$^{+0.11}_{-0.11}$ & -0.97$^{+0.15}_{-0.14}$ & -1.28 &     \\
XBSJ205829.9$-$423634 & 0.232 & 3.62$^{+0.045}_{-0.046}$ & --$^{ }_{ }$ & 
43.44$^{+0.18}_{-0.22}$ & 43.64$^{+0.17}_{-0.22}$ &
7.88$^{+0.12}_{-0.15}$ & --$^{ }_{ }$ & 
7.88$^{+0.12}_{-0.15}$ &  8.02 & -1.15$^{+0.06}_{-0.06}$ & -1.39$^{+0.13}_{-0.16}$ & -1.53 & 1   \\
XBSJ210325.4$-$112011 & 0.720 & --$^{ }_{ }$ & 4.00$^{+0.011}_{-0.011}$ & 
45.48$^{+0.56}_{-0.44}$ & 45.70$^{+0.56}_{-0.44}$ &
--$^{ }_{ }$ & 9.79$^{+0.28}_{-0.22}$ & 
9.79$^{+0.28}_{-0.22}$ &  9.88 & 0.53$^{+0.54}_{-0.40}$ & -1.62$^{+0.61}_{-0.46}$ & -1.71 &   2 \\
XBSJ210355.3$-$121858 & 0.792 & --$^{ }_{ }$ & 3.94$^{+0.021}_{-0.022}$ & 
44.97$^{+0.15}_{-0.22}$ & 45.10$^{+0.15}_{-0.22}$ &
--$^{ }_{ }$ & 9.30$^{+0.09}_{-0.12}$ & 
9.30$^{+0.09}_{-0.12}$ &  9.36 & -0.14$^{+0.11}_{-0.15}$ & -1.80$^{+0.14}_{-0.19}$ & -1.86 &     \\
XBSJ213002.3$-$153414 & 0.562 & 3.36$^{+0.067}_{-0.089}$ & 3.36$^{+0.011}_{-0.011}$ & 
45.61$^{+0.15}_{-0.16}$ & 45.74$^{+0.15}_{-0.15}$ &
8.44$^{+0.14}_{-0.17}$ & 8.53$^{+0.08}_{-0.07}$ & 
8.53$^{+0.08}_{-0.07}$ &  9.13 & 0.39$^{+0.14}_{-0.13}$ & -0.50$^{+0.16}_{-0.15}$ & -1.10 &     \\
XBSJ213729.7$-$423601 & 0.664 & 3.56$^{+0.059}_{-0.033}$ & 3.68$^{+0.014}_{-0.014}$ & 
44.77$^{+0.18}_{-0.15}$ & 44.90$^{+0.18}_{-0.15}$ &
8.41$^{+0.14}_{-0.10}$ & 8.66$^{+0.09}_{-0.08}$ & 
8.41$^{+0.14}_{-0.10}$ &  8.66 & -0.32$^{+0.13}_{-0.10}$ & -1.09$^{+0.19}_{-0.14}$ & -1.34 &     \\
XBSJ213733.2$-$434800 & 0.427 & 3.56$^{+0.084}_{-0.089}$ & 3.60$^{+0.047}_{-0.053}$ & 
44.23$^{+0.20}_{-0.22}$ & 44.42$^{+0.21}_{-0.22}$ &
8.15$^{+0.19}_{-0.20}$ & 8.21$^{+0.14}_{-0.15}$ & 
8.15$^{+0.19}_{-0.20}$ &  8.33 & -0.76$^{+0.16}_{-0.17}$ & -1.27$^{+0.25}_{-0.26}$ & -1.45 &   2 \\
XBSJ214041.4$-$234720 & 0.490 & 4.36$^{+0.083}_{-0.070}$ & 3.93$^{+0.015}_{-0.016}$ & 
44.90$^{+0.12}_{-0.09}$ & 45.13$^{+0.11}_{-0.10}$ &
10.08$^{+0.18}_{-0.14}$ & 9.31$^{+0.06}_{-0.06}$ & 
9.31$^{+0.06}_{-0.06}$ &  9.39 & 0.01$^{+0.10}_{-0.08}$ & -1.66$^{+0.12}_{-0.10}$ & -1.74 &     \\
\hline
\end{tabular}
\end{table}
\end{landscape}
\newpage
\begin{landscape}
\addtocounter{table}{-1}
\begin{table}
\caption{continue}
\begin{tabular}{r r r r r r r r r r r r r r}
\hline\hline
name & z & LogFWHM & LogFWHM  & log$\lambda$L$_{\lambda}$ & Log$\lambda$L$_{\lambda}$ & LogM$_{BH}$ & LogM$_{BH}$ & logM$_{BH}$ & log$M_{BH}$ & Log$\dot{M}$ & LogL/L$_{Edd}$ & LogL/L$_{Edd}$& Flag\\
     &   & H$\beta$ & MgII$\lambda$2798\AA & 5100\AA & 3000\AA& H$\beta$&MgII$\lambda$2798\AA  & best & p$_{rad}$  & & & p$_{rad}$ & \\
(1) & (2) & (3) & (4) & (5) & (6) & (7) & (8) & (9) & (10) & (11) & (12) & (13) & (14) \\
\hline\hline
XBSJ220446.8$-$014535 & 0.540 & 3.73$^{+0.148}_{-0.162}$ & 3.99$^{+0.074}_{-0.089}$ & 
44.12$^{+0.20}_{-0.30}$ & 44.38$^{+0.21}_{-0.30}$ &
8.42$^{+0.31}_{-0.34}$ & 8.96$^{+0.18}_{-0.23}$ & 
8.42$^{+0.31}_{-0.34}$ &  8.64 & -0.38$^{+0.16}_{-0.20}$ & -1.16$^{+0.35}_{-0.39}$ & -1.38 &   2 \\
XBSJ221623.3$-$174317 & 0.754 & --$^{ }_{ }$ & 3.39$^{+0.069}_{-0.083}$ & 
44.70$^{+0.16}_{-0.26}$ & 44.90$^{+0.16}_{-0.27}$ &
--$^{ }_{ }$ & 8.08$^{+0.16}_{-0.20}$ & 
8.08$^{+0.16}_{-0.20}$ &  8.55 & -0.24$^{+0.12}_{-0.16}$ & -0.68$^{+0.20}_{-0.26}$ & -1.16 &     \\
XBSJ223547.9$-$255836 & 0.304 & 3.61$^{+0.034}_{-0.034}$ & 3.39$^{+0.124}_{-0.138}$ & 
43.92$^{+0.15}_{-0.15}$ & 44.15$^{+0.15}_{-0.15}$ &
8.09$^{+0.09}_{-0.10}$ & 7.61$^{+0.23}_{-0.25}$ & 
8.09$^{+0.09}_{-0.10}$ &  8.23 & -0.95$^{+0.12}_{-0.11}$ & -1.40$^{+0.15}_{-0.15}$ & -1.54 &     \\
XBSJ223555.0$-$255833 & 1.800 & --$^{ }_{ }$ & 3.71$^{+0.028}_{-0.030}$ & 
46.23$^{+0.21}_{-0.22}$ & 46.43$^{+0.20}_{-0.22}$ &
--$^{ }_{ }$ & 9.66$^{+0.11}_{-0.13}$ & 
9.66$^{+0.11}_{-0.13}$ & 10.03 & 1.17$^{+0.19}_{-0.20}$ & -0.85$^{+0.22}_{-0.24}$ & -1.22 &     \\
XBSJ223949.8+080926 & 1.406 & --$^{ }_{ }$ & 3.73$^{+0.037}_{-0.041}$ & 
46.01$^{+0.23}_{-0.40}$ & 46.20$^{+0.23}_{-0.39}$ &
--$^{ }_{ }$ & 9.57$^{+0.14}_{-0.21}$ & 
9.57$^{+0.14}_{-0.21}$ &  9.87 & 0.95$^{+0.21}_{-0.35}$ & -0.98$^{+0.25}_{-0.41}$ & -1.28 &     \\
XBSJ224756.6$-$642721 & 0.598 & 3.68$^{+0.035}_{-0.031}$ & 3.67$^{+0.004}_{-0.004}$ & 
45.00$^{+0.11}_{-0.16}$ & 45.23$^{+0.11}_{-0.16}$ &
8.78$^{+0.09}_{-0.10}$ & 8.84$^{+0.06}_{-0.08}$ & 
8.78$^{+0.09}_{-0.10}$ &  9.03 & 0.06$^{+0.11}_{-0.14}$ & -1.08$^{+0.14}_{-0.17}$ & -1.33 &   2 \\
XBSJ225025.1$-$643225 & 1.206 & --$^{ }_{ }$ & 3.76$^{+0.066}_{-0.078}$ & 
45.28$^{+0.18}_{-0.15}$ & 45.48$^{+0.18}_{-0.16}$ &
--$^{ }_{ }$ & 9.17$^{+0.16}_{-0.17}$ & 
9.17$^{+0.16}_{-0.17}$ &  9.36 & 0.28$^{+0.15}_{-0.11}$ & -1.25$^{+0.22}_{-0.20}$ & -1.44 &     \\
XBSJ225050.2$-$642900 & 1.251 & --$^{ }_{ }$ & 3.88$^{+0.036}_{-0.020}$ & 
45.82$^{+0.15}_{-0.16}$ & 45.95$^{+0.15}_{-0.15}$ &
--$^{ }_{ }$ & 9.71$^{+0.11}_{-0.08}$ & 
9.71$^{+0.11}_{-0.08}$ &  9.85 & 0.69$^{+0.11}_{-0.10}$ & -1.38$^{+0.16}_{-0.13}$ & -1.53 &     \\
XBSJ225118.0$-$175951 & 0.172 & 3.40$^{+0.028}_{-0.030}$ & --$^{ }_{ }$ & 
44.29$^{+0.54}_{-0.34}$ & 44.55$^{+0.54}_{-0.34}$ &
7.86$^{+0.27}_{-0.18}$ & --$^{ }_{ }$ & 
7.86$^{+0.27}_{-0.18}$ &  8.43 & -0.33$^{+0.53}_{-0.34}$ & -0.55$^{+0.59}_{-0.38}$ & -1.12 &     \\
XBSJ230400.4$-$083755 & 0.411 & 3.95$^{+0.053}_{-0.048}$ & 3.74$^{+0.013}_{-0.014}$ & 
44.50$^{+0.18}_{-0.15}$ & 44.75$^{+0.17}_{-0.16}$ &
9.05$^{+0.14}_{-0.12}$ & 8.68$^{+0.08}_{-0.09}$ & 
9.05$^{+0.14}_{-0.12}$ &  9.12 & -0.33$^{+0.16}_{-0.15}$ & -1.74$^{+0.21}_{-0.19}$ & -1.81 &     \\
XBSJ230443.8+121636 & 1.405 & --$^{ }_{ }$ & 3.89$^{+0.049}_{-0.056}$ & 
45.71$^{+0.32}_{-0.44}$ & 45.90$^{+0.32}_{-0.44}$ &
--$^{ }_{ }$ & 9.70$^{+0.18}_{-0.25}$ & 
9.70$^{+0.18}_{-0.25}$ &  9.85 & 0.70$^{+0.28}_{-0.30}$ & -1.36$^{+0.33}_{-0.39}$ & -1.51 &   2 \\
XBSJ230459.6+121205 & 0.560 & 3.21$^{+0.063}_{-0.077}$ & --$^{ }_{ }$ & 
44.16$^{+0.19}_{-0.27}$ & 44.29$^{+0.19}_{-0.27}$ &
7.41$^{+0.14}_{-0.18}$ & --$^{ }_{ }$ & 
7.41$^{+0.14}_{-0.18}$ &  8.16 & -0.54$^{+0.06}_{-0.06}$ & -0.31$^{+0.15}_{-0.19}$ & -1.06 & 1 2 \\
XBSJ231342.5$-$423210 & 0.973 & --$^{ }_{ }$ & 3.74$^{+0.046}_{-0.052}$ & 
45.24$^{+0.12}_{-0.10}$ & 45.44$^{+0.11}_{-0.10}$ &
--$^{ }_{ }$ & 9.12$^{+0.11}_{-0.11}$ & 
9.12$^{+0.11}_{-0.11}$ &  9.33 & 0.30$^{+0.08}_{-0.06}$ & -1.18$^{+0.14}_{-0.13}$ & -1.39 &     \\
XBSJ231601.7$-$424038 & 0.383 & 3.66$^{+0.033}_{-0.057}$ & --$^{ }_{ }$ & 
44.05$^{+0.11}_{-0.16}$ & 44.32$^{+0.11}_{-0.16}$ &
8.25$^{+0.09}_{-0.14}$ & --$^{ }_{ }$ & 
8.25$^{+0.09}_{-0.14}$ &  8.48 & -0.52$^{+0.10}_{-0.13}$ & -1.13$^{+0.13}_{-0.19}$ & -1.36 &     \\
\hline
\end{tabular}
Column~1: source name; 
Column~2: redshift; 
Column~3: Logarithm  of the 
FWHM of the broad component of the H$\beta$ (in km s$^{-1}$); 
Column~4: Logarithm  of the 
FWHM of the broad component of the MgII$\lambda$2798\AA\ (in km s$^{-1}$); 
Column~5: Logarithm of the monochormatic luminosity at 5100\AA\ (taken from
the SED fitting presented in Marchese et al. (2012);
Column~6: Logarithm of the monochormatic luminosity at 3000\AA\ (taken from
the SED fitting presented in Marchese et al. (2012);
Column~7: Logarithm of the black hole mass derived from H$\beta$ (in solar mass 
units);
Column~8: Logarithm of the black hole mass derived from MgII$\lambda$2798\AA\ 
(in solar mass units);
Column~9: Logarithm of the black hole mass considered as best estimate 
(in solar mass units);
Column~10: Logarithm of the best estimate black hole mass corrected for the effect of radiation pressure
(in solar mass units);
Column~11: Logarithm of the absolute accretion rate (in units of solar masses
per year);
Column~12: Logarithm of the Eddington ratio;
Column~13: Logarithm of the Eddington ratio corrected for the effects of 
radiation pressure;
Column~14: Flag indicating uncertainity in the best estimate mass 
(1=problems during the spectral fitting procedure, 2=low S/N ($<$5) in the spectral region of the 
line used for the mass estimate)

(Errors are at 68\% confidence level)
\end{table}
\end{landscape}
\end{document}